\documentclass[12pt]{article}

\usepackage{amsmath}
\usepackage{amstext}
\usepackage{amsthm}
\usepackage{amssymb}
\usepackage{graphicx}
\usepackage{mathrsfs}
\usepackage[bottom]{footmisc}
\usepackage{indentfirst}
\usepackage{upref}
\usepackage{cite}
\usepackage{units}
\newcommand{\be}{\begin{equation}}
\newcommand{\ee}{\end{equation}}
\newcommand{\ba}{\begin{eqnarray}}
\newcommand{\ea}{\end{eqnarray}}
\newcommand{\bc}{\begin{center}}
\newcommand{\ec}{\end{center}}

\newcommand{\bay}{\begin{array}{rcl}}
\newcommand{\eay}{\end{array}}

\def\ie{{i.e. }}
\def\lp{\ell_{\rm Pl}}
\def\LH{\ell_{\rm H}}

\def\tr{t_{\rm tr}}
\def\mp{m_{\rm Pl}}
\def\tp{t_{\rm Pl}}
\def\OM{\Omega_{\rm M}}
\def\OL{\Omega_{\Lambda}}

\def\lst{{\lambda}_\ast}
\def\laz{\lambda_{\rm za}}
\def\gst{{g}_\ast}

\def\OLS{\Omega_\Lambda^\ast}
\def\OMS{\Omega_{\rm M}^\ast}
\def\barg{\bar{G}}
\def\lat{\lambda_{\rm T}}
\def\ett{\eta_{\rm N}}
\def\kat{k_{\rm T}}
\def\gat{g_{\rm T}}
\def\hT{H_{\rm T}}
\def\h0{H_{0}}
\def\aT{a_{\rm T}}
\def\gs{g_{\rm s}}
\def\ls{\lambda_{\rm s}}
\def\TT{T_{\rm T}}
\def\tT{t_{\rm T}}
\def\unm{\frac{1}{2}}
\def\calp{\widetilde{\mathscr P}} 
\def\clp{{\mathscr P}} 
\def\clm{{\cal M}} 
\def\UU{{\cal U}} 
\def\ssc{S_{\rm c}} 
% ... 

\begin{document}
\title{{\bf Entropy signature of the running cosmological constant}
  \raisebox{3cm}[0mm][0mm]{ 
\begin{minipage}[b]{0cm}
  \normalsize \noindent \mbox{\hspace{-0.5cm}
        MZ-TH/07-10
  }
\end{minipage}}}
\date{}
\maketitle
\vspace{-1cm}
\begin{center}
{\sc A. Bonanno$^{1,2}$ and M. Reuter$^3$ }\\
\vspace{1cm}
{\it $^1$INAF-Catania Astrophysical Observatory,
Via S.Sofia 78, 95123 Catania, Italy\\
$^2$INFN, Via S.Sofia 64, 95123 Catania, Italy, E-mail: abo@oact.inaf.it}\\
  { \it $^3$Institute of Physics, University of Mainz, Staudingerweg 7,
    D-55099 Mainz, Germany, E-mail: reuter@thep.physik.uni-mainz.de}
\end{center}

\vspace{.5cm}
\thispagestyle{empty}
\begin{abstract}
Renormalization group (RG) improved cosmologies based upon a RG trajectory
of Quantum Einstein Gravity (QEG) with realistic parameter values are investigated
using  a system of cosmological evolution equations which allows for an 
unrestricted energy exchange between the vacuum and the matter sector. It is 
demonstrated that the scale dependence of the gravitational parameters, the
cosmological constant in particular, leads to an entropy production in the matter system.
The picture emerges that the Universe started out from a state of vanishing entropy,
and that the radiation entropy observed today is essentially due to the  
coarse graining (RG flow) in the quantum gravity sector which is related to the expansion of the Universe. Furthermore, the RG improved
field equations are shown to possess solutions with an epoch of power law inflation immediately
after the initial singularity. The inflation is driven by the cosmological constant and ends 
automatically once the RG running has reduced the vacuum energy to the level of the matter
energy density.
\end{abstract}

\newpage
\section{Introduction}
\renewcommand{\theequation}{1.\arabic{equation}}
\label{3intro}
\setcounter{equation}{0}
After the introduction of a functional renormalization group for gravity
\cite{mr} detailed investigations of the nonperturbative renormalization
group (RG) behavior of 
Quantum Einstein Gravity have become possible 
\cite{mr,percadou,oliver1,frank1,oliver2,oliver3,oliver4,souma,perper1,
codello,frank2,litimgrav,prop,blaub,essential,hier}.
The exact RG equation underlying this approach defines a Wilsonian
RG flow on a theory space which consists of all diffeomorphism invariant
functionals of the metric $g_{\mu\nu}$. The approach   turned out
to be an ideal setting for investigating the asymptotic safety scenario in gravity
\cite{wein,livrev} and, in fact, substantial evidence was found for the
nonperturbative renormalizability of Quantum Einstein Gravity.
The theory emerging  from this construction (sometimes denoted ``QEG")
is not a quantization of classical general relativity. Instead, its bare action
corresponds to nontrivial fixed point of the RG flow and is 
a {\it prediction} therefore, and not as usually in quantum field theory an 
ad hoc assumption defining some ``model". Independent support
for the asymptotic safety conjecture came from a 2-dimensional symmetry reduction of
the gravitational path-integral \cite{max}. The approach of \cite{mr} employs the effective 
average action  \cite{avact,ym,avactrev}
which has crucial advantages as compared to other continuum implementations of the 
Wilson RG \cite{bagber}, in particular it is closely related to the standard 
effective action and defines a family of effective field theories
$\{ \Gamma_k[g_{\mu\nu}], 0 \leq k < \infty \}$ labeled by the coarse graining scale $k$.
The latter property opens the door to a rather direct extraction of physical
information from the RG flow, at least in single-scale cases: If the physical process 
or phenomenon under consideration involves only a single typical momentum
scale $p_0$ it can be described by a tree-level evaluation of $\Gamma_k[g_{\mu\nu}]$, with $k=p_0$.
The precision which can be achieved by this effective field theory description
depends on the size of the fluctuations relative to mean values. If they are large, or if more than
one scale is involved, it might be necessary to go be beyond the tree analysis. 

The effective field theory techniques proved useful for an understanding 
of the scale dependent geometry of the effective QEG spacetimes
\cite{oliverfrac,jan1,jan2}. In particular it has been shown \cite{oliver1,oliver2,oliverfrac}
that these spacetimes have fractal properties, with  a fractal dimension of 2 at small,
and 4 at large distances. The same dynamical dimensional reduction was also observed
in numerical studies of Lorentzian dynamical triangulations 
\cite{ajl1,ajl2,ajl34} and in \cite{ncgeom} A.Connes et al. speculated about its
possible relevance to the noncommutative geometry of the standard model.

The RG flow of the effective average action, obtained by different truncations of theory space, 
has been the basis of various investigations of ``RG improved" black hole and cosmological
spacetimes [31-41].
We shall discuss some aspects of this method below.

A special class of RG trajectories obtained from QEG in the Einstein-Hilbert approximation
\cite{mr}, namely those of the ``Type IIIa'' \cite{frank1}, possess all the qualitative
properties one would expect from the  RG trajectory describing gravitational phenomena
in the real Universe we live in. In particular they can have a long classical regime and a small,
positive cosmological constant in the infrared (IR). Determining its parameters from observations,
one finds \cite{h3} that,  according to this particular QEG trajectory, the running cosmological
constant $\Lambda(k)$ changes by about 120 orders of magnitude between $k$-values of the order
of the Planck mass and macroscopic scales, while the running Newton constant $G(k)$ has no
strong $k$-dependence in this regime. For $k> \mp$, the non-Gaussian fixed point (NGFP)
which is responsible for the renormalizability of QEG controls their scale dependence. In the deep
ultraviolet $(k\rightarrow \infty)$, $\Lambda(k)$ diverges and $G(k)$ approaches zero.

In the present paper we are going to ask whether there is any experimental or observational 
evidence that would  hint at this enormous scale dependence of the gravitational parameters,
the cosmological constant in particular. Clearly the natural place to search for such phenomena
is cosmology. Even though it is always difficult to give a precise physical interpretation
to the RG scale $k$ is is fairly certain that any sensible identification of $k$ in terms
of cosmological quantities will lead to a $k$ which decreases during the expansion
of the Universe. As a consequence, $\Lambda(k)$ will also decreases as the Universe expands.
Already the purely qualitative assumption of a {\it positive} and {\it decreasing}
cosmological constant supplies an interesting hint as to which 
phenomena might reflect a possible $\Lambda$-running. 
\begin{figure}[t]
\begin{center}
\includegraphics[width=0.5\textwidth]{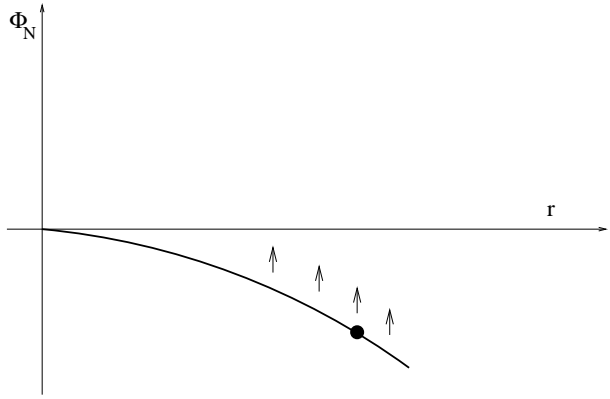}
\end{center}
\caption{ The quasi-Newtonian potential corresponding to de Sitter space.
The curve moves upward as the cosmological constant decreases.} 
\label{fig1}
\end{figure}

To make the argument as simple as possible, let us first consider a Universe without matter,
but with a positive $\Lambda$. Assuming maximal symmetry, this is nothing but  de Sitter
space, of course. In static coordinates its metric is 
\be\label{1.1}
ds^2 = -\Big (1+2\Phi_{\rm N}(r) \Big) dt^2+\Big (1+2\Phi_{\rm N}(r)\Big)^{-1}dr^2 +
r^2 (d\theta^2 +\sin^2 \theta d\phi^2)
\ee
with 
\be\label{1.2}
\Phi_{\rm N}(r) = -\frac{1}{6}\; \Lambda\; r^2
\ee
In the weak field and slow motion limit $\Phi_{\rm N}$ has the interpretation
of a Newtonian potential, with a correspondingly simple physical interpretation.
Fig.~\ref{fig1} shows $\Phi_{\rm N}$ as a function of $r$; for $\Lambda >0$ it is 
an upside-down parabola. Point particles in this spacetime, symbolized by the 
black dot in Fig.~\ref{fig1}, ``roll down the hill'' and are rapidly driven away
from the origin and from any other particle. Now assume that the magnitude of $\Lambda$
is slowly (``adiabatically'') decreased. This will cause the potential $\Phi_{\rm N}(r)$ 
to move upward as a whole, its slope decreases. So the change in $\Lambda$ increases the particle's
potential energy.  This is the simplest way of understanding that a {\it positive decreasing}
cosmological constant has the effect of ``pumping'' energy into the matter degrees of freedom.
More realistically one will describe the matter system in a hydrodynamics or quantum field
theory language and one will include its backreaction onto the metric. But the basic conclusion, 
namely that a slow decrease of a positive $\Lambda$ transfers energy into the matter system, will
remain true. 

We are thus led to the suspect that, because of the decreasing cosmological constant, 
there is a continuous inflow of energy into the cosmological fluid contained in an 
expanding Universe. It will ``heat up'' the fluid or, more exactly, lead to a slower decrease
of the temperature than in standard cosmology. Furthermore, by elementary thermodynamics, it will 
{\it increase} the entropy of the fluid. If during the time $dt$ an amount of heat
$d Q>0$ is transferred into a volume $V$ at the temperature $T$ the entropy changes by an amount $dS=dQ/T>0$.
To be as conservative (i.e., close to standard cosmology) as possible, we assume that this process
is reversible. If not, $dS$ is even larger.

In standard Friedmann-Robertson-Walker (FRW) cosmology the expansion is adiabatic,
the entropy (within a comoving volume) is constant. It has always been somewhat puzzling therefore
where the huge amount of entropy contained in the present Universe comes from. 
Presumably it is dominated by the CMBR photons which  contribute an amount
of about $10^{88}$ to the entropy within the present Hubble sphere. (We use units such that
$k_{\rm B}=1$. ) In fact, if it is really true that no entropy is produced during the 
expansion then the Universe would have had an entropy of at least $10^{88}$ immediately 
after the initial singularity which for various reasons seems quite unnatural
\cite{bigbangentropy}. In scenarios which invoke a ``tunneling from nothing'', for instance,
spacetime was ``born'' in a pure quantum state, so the very early Universe is expected to have 
essentially no entropy \cite{tunnel}. Usually it is argued that the entropy present today is the result
of some sort of ``coarse graining'' which, however, typically is not considered an active part of the
cosmological dynamics in the sense that it would have an impact on the 
time evolution of the metric, say.

In the present paper we are going to argue that in principle the entire entropy of the massless fields in the present
universe can be understood as arising from the mechanism described above, the ``heating'' of 
matter by a decreasing cosmological constant. If energy can be exchanged freely between
the cosmological constant and the matter degrees of freedom, the entropy observed today 
is obtained precisely if the initial entropy at the ``big bang'' vanishes.

The assumption that the matter system must allow for an unhindered energy exchange with
$\Lambda$ is nontrivial. In refs. \cite{cosmo1} and \cite{cosmofrank}, henceforth referred as
[I] and [II], respectively, ``RG improved'' cosmologies were studied which, too, are based upon
the RG trajectories of QEG. In these investigations it has been assumed, however,
that there is no injection of energy into the matter system due to the time dependence of
$\Lambda$, and that the evolution is adiabatic therefore. In the present paper we explore the 
opposite situation of a completely unobstructed energy transfer. Technically this 
amounts to dropping the so called ``consistency condition'' imposed in [I] and [II].
Which one of the two cases is more realistic depends on the cosmological epoch and on 
properties of the matter model (particle masses, couplings, etc.).

As in [I] and [II] the computational setting of the present paper are the RG
improved Einstein equations: By means of a suitable cutoff identification $k=k(t)$
we turn the scale dependence of $G(k)$ and $\Lambda(k)$ into a time dependence, and then substitute the resulting $G(t)\equiv G(k(t))$
and $\Lambda(t)\equiv \Lambda(k(t))$ into the Einstein equations. We shall obtain the RG trajectory
by solving the flow equation for the Einstein-Hilbert truncation with a
sharp cutoff \cite{mr,frank1}. We then construct quantum corrected cosmologies 
by (numerically) solving   the RG improved cosmological evolution equations. 

We model the matter in the early Universe by a gas with $n_{\rm b}$ 
bosonic and $n_{\rm f}$ fermionic massless degrees of freedom, all at  the same
temperature. {\it In equilibrium} its energy density, pressure, and entropy density
are given by the usual relations ($n_{\rm eff}=n_{\rm b}+\frac{7}{8}n_{\rm f}$)
\begin{subequations}\label{1.3}
\ba\label{1.3a}
&&\rho = 3\; p = \frac{\pi^2}{30} \; n_{\rm eff} \; T^4\\[2mm] 
&&s = \frac{2\pi^2}{45}\; n_{\rm eff} \; T^3\label{1.3b}\\[2mm]
&&\text{ \hspace{-6.6cm}  so that in terms of $U\equiv \rho \; V$ and $S\equiv s\; V$,} \nonumber\\[2mm]
&&T\; dS = dU + p\; dV\label{1.3c}
\ea
\end{subequations}
In an out-of-equilibrium process of entropy generation the question arises 
how the various thermodynamical quantities are related then. To be as conservative as possible,
we make the assumption that the irreversible inflow of energy
destroys thermal equilibrium as little as possible in the sense that the equilibrium
relation (\ref{1.3}) continue to be (approximately) valid.

This kind of thermodynamics in an FRW-type cosmology with a decaying cosmological constant
has been analyzed in detail by Lima \cite{lima1}, see also \cite{lima2}. It was shown that if
the process of matter creation $\Lambda(t)$ gives rise to is such that the specific entropy
per particle is constant, the relations of equilibrium thermodynamics are preserved.
This means that no finite thermalization time is required since the particles originating from 
the decaying vacuum are created in equilibrium with the already existing ones. Under these
conditions it is also possible to derive a generalized black body spectrum which is
conserved under time evolution. Such minimally non-adiabatic processes
were termed ``adiabatic'' (with the quotation marks) in refs. \cite{lima1,lima2}.

In section 3 of the present paper we shall discuss the ``adiabatic'' generation of entropy in the
framework of the RG improved cosmology.

\vspace{5mm}
There is another, more direct potential consequence of a decreasing positive cosmological
constant which we shall also explore in this paper, namely a period of automatic
inflation during the very first stages of the cosmological evolution. It is not surprising,
of course, that a positive $\Lambda$ can cause an accelerated expansion, but in the
classical context the problem with a $\Lambda$-driven inflation is that it would never
terminate once it has started. In popular models of scalar driven inflation \cite{lily}
this problem is circumvented by designing the inflaton potential in such a way that it
gives rise to a vanishing vacuum energy after a period of ``slow roll''. 

In this paper we shall see that generic RG cosmologies based upon the QEG trajectories  
have an era of $\Lambda$-driven inflation immediately after the big bang which ends automatically as a 
consequence of the RG running of $\Lambda(t)$. Once the scale $k$ drops significantly below 
$\mp$, the accelerated expansion ends because the vacuum energy density $\rho_\Lambda$ is already
too small to compete with the matter density. Clearly this is a very attractive scenario:
neither to trigger inflation nor to stop it one needs any ad hoc ingredients such
as an inflaton field or a special potential. It suffices to include the leading quantum
effects in the gravity + matter system.

It is to be emphasized that the present investigations are not some sort of ``model
building'' of decaying-$\Lambda$ cosmologies; they rather deal with consequences of the 
{\it computable} scale dependence of $\Lambda$ and $G$. Besides the validity
of the mean field description and the above assumptions about the thermodynamical properties
of matter, the only other assumption we make is that the renormalization effects of the matter
fields, which are not taken into account explicitly, do not alter the RG flow of pure gravity as far as qualitative
features and orders of magnitude are concerned. If so, it makes sense to confront the RG trajectories
of pure QEG with observations in the real world.

\vspace{5mm}
Let us briefly review how the type IIIa trajectories of the Einstein-Hilbert truncation
can be matched against the observational data \cite{bh}. This analysis is fairly robust 
and clearcut; it does not involve the NGFP. All that is needed is the RG flow linearized 
about the GFP. It reads \cite{mr}
\ba
&&\Lambda(k) = \Lambda_0 + \nu \; \barg \, k^4+\cdots\\
&&G(k) =\barg+\cdots\nonumber
\ea
Or, in terms of the dimensionless couplings $g(k)\equiv k^2 G(k)$, and
$\lambda(k) \equiv \Lambda(k)/k^2$:
\ba\label{1.11}
&&\lambda(k) = \Lambda_0/k^2 + \nu \; \barg \, k^2+\cdots\\
&&g(k) =\barg \; k^2+\cdots\nonumber
\ea
In the linear  regime of the GFP, $\Lambda$ displays a running $\propto k^4$
which is seen in perturbation theory already,  
and $G$ is approximately constant. Here $\nu$ is a positive 
constant of order unity \cite{mr,frank1}, 
\be\label{1.12}
\nu \equiv \frac{1}{4\pi} \Phi^{1}_{2}(0)\equiv \frac{\varphi_2}{4\pi}
\ee 
Eqs.(\ref{1.11}) are valid if $\lambda(k) \ll 1$ and $g(k)\ll 1$. They
describe a 2-parameter family of RG trajectories labeled by the pair $(\Lambda_0, \barg)$.
It will prove convenient to use an alternative labeling $(\lat, \kat)$
with 
\ba\label{1.13}
&&\lat \equiv (4 \; \nu \; \Lambda_0 \; \barg)^{1/2} \\
&&\kat \equiv  \Big ( \frac{\Lambda_0}{\nu \; \barg} \Big )^{1/4} \nonumber
\ea
The old labels are expressed in terms of the new ones as
\ba\label{1.14}
&&\Lambda_0 = \frac{1}{2} \lat \; \kat^2\\
&&\barg = \frac{\lat}{2\,\nu\,\kat^2}\nonumber
\ea
It us furthermore convenient to introduce the abbreviation (not an independent
label)
\be\label{1.15}
\gat\equiv \frac{\lat}{2\,\nu}\equiv \frac{\lat}{(\varphi_2/2\pi)}
\ee
When parameterized by the pair $(\lat,\kat)$ the trajectories assume the form
\ba\label{1.16}
&&\Lambda(k) = \frac{1}{2} \; \lat\; \kat^2 \; \Big [ 1+(k/\kat)^4\Big ]
\equiv \Lambda_0 \Big [ 1+(k/\kat)^4 \Big ]\\[2mm]
&&G(k) = \frac{\lat}{2\,\nu\,\kat^2}\equiv \frac{\gat}{\kat^2}\nonumber
\ea
or, in dimensionless form, 
\ba\label{1.17}
&&\lambda(k) = \frac{1}{2}\; \lat \; \Big [ \Big (\frac{\kat}{k}\Big)^2+ 
\Big ( \frac{k}{\kat} \Big )^2 \Big ]\\[2mm]
&&g(k) = \gat \, \Big ( \frac{k}{\kat} \Big )^2\nonumber
\ea
Note that $\lambda(k)$ is invariant under the ``duality transformation" 
\cite{jan1} $k\mapsto \kat^2/k$. As for the interpretation of the new variables, 
it is clear that $\lat \equiv \lambda(k\equiv \kat)$ and $\gat\equiv g(k=\kat)$,
while $\kat$ is the scale at which $\beta_\lambda$ (but not $\beta_g$) vanishes according
to the linearized running (\ref{1.17}):
\be\label{1.18}
\beta_\lambda(\kat)\equiv k\frac{d \lambda(k)}{dk} \Big |_{k=\kat} =0
\ee
Thus we see that $(\gat,\lat)$ are the coordinates of the turning point T
of the type IIIa trajectory considered, and $\kat$ is the scale at which it is
passed. It is convenient to refer the ``RG time'' $\tau$ to this scale:
\be\label{1.19}
\tau(k) \equiv \ln (k/\kat)
\ee
so that $\tau>0$ ($\tau<0$) corresponds to the ``UV regime'' (``IR regime'') where
$k>\kat$ ($k<\kat)$. In terms of the RG time, 
\ba\label{1.20}
&&\lambda(\tau) = \lat \; \cosh (2\tau)\\[2mm]
&&g(\tau) = \gat \; \exp (2\tau)\nonumber
\ea

Let us now hypothesize that, within a certain range of $k$-values, the RG trajectory
realized in Nature can be approximated by (\ref{1.17}). In order to determine
its parameters $(\Lambda_0, \barg)$ or $(\lat, \kat)$ we must perform a measurement
of $G$ and $\Lambda$. If we interpret the observed values
\ba\label{1.21}
&&G_{\rm observed} = \mp^{-2}, \;\;\;\;\;\; \mp\approx 1.2\times 10^{19} \, {\rm GeV}\\[2mm]
&&\Lambda_{\rm observed} = 3\,\Omega_{\Lambda 0}\,H_0^2\approx 10^{-120}\, \mp^2\nonumber
\ea
as the running $G(k)$ and $\Lambda(k)$ evaluated at a scale $k\ll \kat$, then we get from
(\ref{1.16}) that $\Lambda_0 =\Lambda_{\rm observed} $ and 
$\barg = G_{\rm observed}$. Using (\ref{1.13}) and $\nu = O(1)$ this leads to the 
order-of-magnitude estimates
\ba\label{1.22}
&&\gat\approx \lat \approx 10^{-60}\\[2mm]
&&\kat\approx 10^{-30}\;\mp\approx (10^{-3} {\rm cm})^{-1}\nonumber
\ea
Because
of the tiny values of $\gat$ and $\lat$ the turning point lies in the linear regime of GFP. 

Up to this point we discussed only that segment of the ``trajectory realized in Nature'' which lies
inside the linear regime of the GFP. The complete RG trajectory obtains by continuing
this segment with the flow equation both into the IR and into the UV, 
where it ultimately spirals into the 
NGFP. While the UV-continuation is possible within the Einstein-Hilbert
truncation, this approximation breaks down in the IR when $\lambda(k)$ approaches $1/2$.
A rough estimate for the ``termination" scale $k_{\rm term}$ at which this happens can be obtained from
(\ref{1.11}) for $k\ll \mp$: $\lambda(k) \approx \Lambda_0/k^2=\Lambda_{\rm observed}/k^2=
3\Omega_{\Lambda 0}(H_0/k)^2$. Since the observations show that $\Omega_{\Lambda 0}=O(1)$, 
we have $\lambda(k)=O(1)$ exactly when $k/H_0=O(1)$. Stated differently, it is precisely for scales
of the order of the present Hubble parameter that the Einstein-Hilbert truncation becomes
insufficient. In \cite{h2,h3} it was speculated that close to this regime strong IR
renormalization effects could set in  which perhaps might mimic the presence of dark matter.
Whether these effects actually are there is an open problem; it will not affect the
discussions in the present paper.

Let us try to interpret the trajectory found above in a cosmological context and let us 
ask during which cosmological epoch the Universe a whole passed through the turning point.
Using the natural cutoff identification $k(t)\approx H(t)$ equation (\ref{1.22}) tells
us that this happened when the Hubble parameter was of the order $H_{\rm T}\approx 10^{-30}\mp$.
We can estimate the corresponding redshift $z_{\rm T}$ by exploiting that for 
$k\lesssim \kat$  the impact of the cosmological constant is small so that we have a standard 
radiation dominated FRW cosmology 
with $a(t) \propto t^{1/2} \propto H(t)^{-1/2}$. Neglecting
the comparatively short matter dominated era we can then relate the scale factor at the turning point,
$a_{\rm T}$, to its present value $a_0$ by
\be\label{1.23}
1+z_{\rm T}=\frac{a_0}{a_{\rm T}} = \Big (\frac{H_{\rm T}}{H_0}\Big )^{1/2}\approx 10^{15}
\ee
Here we used that $\hT/H_0 \approx 10^{-30} (\mp/H_0) \approx 10^{30}$. Since the temperature
behaves as $T\propto 1/a$ we can express its value at the turning point in terms
of the present $T_0\approx 2.7 \; \rm K$:
\be\label{1.24}
\TT= 10^{15} \, T_0 \approx 3\times 10^{15} \; \rm K\approx 300 \, {\rm GeV}
\ee
Thus we see that the Universe passed through the turning point at about the time of the
electroweak phase transition. This is a quite remarkable coincidence which might 
have a deeper meaning perhaps. 
It is, however, important to bear in mind how precisely this result is to be interpreted:
At the cosmological time when the electroweak phase transition took place, $t_{\rm EWPT}$,
the Universe {\it on scales of the Hubble radius $1/H(t_{\rm EWPT}$}) is effectively
described by $G(k)$ and $\Lambda(k)$ evaluated at $k\approx k_{\rm T}$. 
If, on the other hand, one wants to describe the {\it microphysics} of the phase transition
where the pertinent scale is the transition temperature $T_{\rm EWPT}=O(100\;  \rm GeV)$
then one should set $k\approx T_{\rm EWPT}$, and this scale is far higher than $k_{\rm T}$. 

The remaining sections of this paper are organized as follows. In section 2 we discuss
the essential properties of the RG improved Einstein equations, and in section 3 we
analyze the mechanism of entropy production they give rise to.
In section 4 we obtain analytical solutions to those equations, valid during specific
cosmological epochs, in particular in the very early Universe whose properties are crucially
determined by the RG fixed point. Then, in section 5, we study the coupled system of RG and 
cosmological evolution equations with numerical methods; we obtain complete cosmological
histories for a RG trajectory with realistic parameter values. The phenomenon of 
automatic inflation in the fixed point regime is discussed in section 6, and section 7
contains the conclusions.
\section{The improved field equations}
\renewcommand{\theequation}{2.\arabic{equation}}
\setcounter{equation}{0}
We assume that $G(k)$ and $\Lambda(k)$ have been converted to functions of the 
cosmological time, $G(t)$ and $\Lambda(t)$, by an appropriate cutoff identification $k=k(t)$
whose precise form is not important for the time being. We then ``RG improve'' the Einstein
equations by substituting these functions for their classical counterparts:
$G_{\mu\nu}=-\Lambda(t)g_{\mu\nu}+8\pi G(t) T_{\mu\nu}$. We specialize $g_{\mu\nu}$
to describe a spatially flat $(K=0)$ Robertson-Walker metric with scale factor $a(t)$,
and we take ${T_\mu}^\nu = {\rm diag}[-\rho,p,p,p]$ to be the energy momentum
tensor of an ideal fluid with equation of state $p=w\rho$ where $w>-1$ is constant.
Then the improved Einstein equation boils down 
to the modified Friedmann equation and a continuity equation:
\begin{subequations}
\ba\label{2.1a}
&&H^2 = \frac{8\pi}{3}\; \barg \; \rho_{\rm eff}\\[2mm]
&&\dot\rho_{\rm eff}+3 \, H (\rho_{\rm eff}+p_{\rm eff})=0 \label{2.1b}
\ea
\end{subequations}
Here $\barg$ is an arbitrary constant,  and 
\ba 
&&\rho_{\rm eff} \equiv \frac{G(t)}{\barg}\; (\rho +\rho_{\Lambda})\\[2mm]
&&p_{\rm eff} \equiv \frac{G(t)}{\barg}\; (p+p_\Lambda)\nonumber
\ea
where 
\be
\rho_\Lambda = -p_\Lambda = \frac{\Lambda(t)}{8\pi G(t)}
\ee
Eqs.(\ref{2.1a}) and (\ref{2.1b})  have the same  appearance as in classical
FRW cosmology with $\Lambda =0$ except that $\rho$ and $p$ are replaced by
$\rho_{\rm eff}$ and $p_{\rm eff}$, respectively. Written more explicitly, this system of equations
reads 
\begin{subequations}
\ba\label{2.5a}
&&H^2 = \frac{8\pi}{3} G(t) \; \rho + \frac{1}{3}\Lambda(t)\\[2mm]
&&\dot\rho+3H(\rho+p)=-\frac{\dot{\Lambda}+8\pi \; \rho \; \dot{G}}{8\pi \; G}\label{2.5b}
\ea
\end{subequations}
The modified continuity equation (\ref{2.5b}) is the integrability condition for the 
improved Einstein equation implied by Bianchi's identity, 
$D^\mu [-\Lambda(t) g_{\mu\nu} + 8\pi G(t) T_{\mu\nu}]=0$. 
It describes the energy exchange 
between the matter and gravitational degrees of freedom (geometry).

In [I] and [II] the special case was considered where 
the coupled dynamics is such that there 
occurs no significant exchange between the two sectors. In this case Eq.(\ref{2.5b}) is 
solved in the form $0=0$, i.e.  both sides vanish separately:
\begin{subequations}
\ba\label{2.6a}
&&\dot\rho+3H(\rho+p)=0\\[2mm]
&&\dot\Lambda +8\pi \; \rho \; \dot {G}=0\label{2.6b}
\ea
\end{subequations}
Eq.(\ref{2.6a}) was referred to as the ``ordinary continuity equation'' and
(\ref{2.6b}) as the ``consistency condition''. Clearly the set of equations
(\ref{2.5a}), (\ref{2.6a}), (\ref{2.6b}) is stronger and more constraining than
(\ref{2.5a}), (\ref{2.5b}). In fact, it is quite nontrivial that the former has physically
acceptable solutions at all. In [I] they were found analytically in the NGFP
regime, and in [II] the complete cosmology was  obtained using  a  special, dynamically
adjusted cutoff identification (otherwise no solution exists).

The analyses in [I] and  [II] dealt with a situation where, for an
unspecified dynamical reason, the exchange of energy and momentum between
matter and the running couplings is completely forbidden. In the following we consider
the other limiting case where this exchange is possible without any 
obstructions\footnote{This is the case which has also been studied in most of the early
phenomenological papers on cosmologies with time dependent $\Lambda$ \cite{mrcwlambda}.}.
We shall analyze the coupled system (\ref{2.5a}), (\ref{2.5b}) where we now accept
any solution of (\ref{2.5b}), not necessarily of the form ``$0=0$''.

For later use let us note that upon
defining the critical density
\be\label{2.60}
\rho_{\rm crit}(t)\equiv \frac{3 \; H(t)^2}{8\pi \; G(t)}
\ee
and the relative densities $\Omega_{\rm M}\equiv \rho/\rho_{\rm crit}$ and 
$\Omega_\Lambda=\rho_\Lambda/\rho_{\rm crit}$ the modified Friedmann equation (\ref{2.5a})
can be written as
\be\label{2.61}
\OM(t) +\OL(t) = 1
\ee
We emphasize that this ``sum rule'' is valid for arbitrary functions $G(t)$ and 
$\Lambda(t)$. (Recall that we consider flat time slices throughout.)
\subsection{An algorithm for generating solutions}
Let $G(t)$ and $\Lambda(t)$ be arbitrary prescribed functions.
If $H(t)\not =0$, the resulting solutions $\Big (  H(t), \rho(t)\Big )$ of the 
system (\ref{2.1a}), (\ref{2.1b}) or equivalently (\ref{2.5a}), (\ref{2.5b})
satisfy the equations 
\begin{subequations}
\ba\label{2.7a}
&&\dot H = - 4\pi (1+w) G(t) \rho\\[2mm]
&&\dot H = -\frac{1}{2}(3+3w)\Big [ H^2 -\frac{1}{3}\Lambda(t)\Big ]\label{2.7b}\\[2mm]
&&\rho = \frac{3}{8\pi G(t)}\Big [ H^2-\frac{1}{3}\Lambda(t)\Big ]\label{2.7c}
\ea
\end{subequations}
This statement is easily proven by differentiating the modified Friedmann equation (\ref{2.1a}).
What is less trivial, but more important from the point of view of finding solutions,
is that its converse is also true. To be precise, we have the following 
algorithm for generating solutions to the improved field equations:

{\it Let $G(t)$ and $\Lambda(t)$ be prescribed functions and $H(t)$ a solution of }
\be\label{hht}
\dot H = -\frac{1}{2}(3+3w)\Big [ H^2-\frac{1}{3} \Lambda(t) \Big ] 
\ee
{\it Let furthermore $\rho(t)$ be defined in terms of this solution according to}
\be\label{pippo}
\rho = \frac{3}{8\pi G(t)} \Big [ H^2 -\frac{1}{3}\Lambda(t) \Big ]
\ee
{\it Then the pair $\Big ( H(t), \rho(t)\Big )$ is a solution to the system of differential equations
}(\ref{2.5a}), (\ref{2.5b}) {\it for the equation of state $p=w\rho$, provided $H(t)\not = 0$.  }

From (\ref{pippo}) it is obvious that the solution generated by the algorithm 
satisfies (\ref{2.5a}). To show that it also satisfies (\ref{2.5b}) one
exploits that, for $H\not = 0$, (\ref{2.5b}) together with (\ref{pippo}) is equivalent
to $2\dot{H}+(3+3w)(H^2-\Lambda/3)=0$, which is nothing but (\ref{hht}).

The existence of this simple algorithm is non-trivial since it amounts to decoupling the 
two differential equations for $\Big (H(t),\rho(t) \Big )$. In fact, one has to solve only a single
differential equation, (\ref{pippo}), involving only one of the external functions, $\Lambda(t)$.
If this has been achieved, $\rho(t)$ is given in terms of $H(t)$ by an explicit formula, and
only here $G(t)$ enters.

For later use let us also note that by using the definition of the deceleration parameter 
\be\label{2.9}
q \equiv -\frac{a\; \ddot{a}}{\dot{a}^2} = -\frac{\dot{H}}{H^2}-1
\ee
together with the differential equation (\ref{pippo}) one obtains a simple expression
for $q$ in terms $\Omega_\Lambda$:
\be\label{2.10}
q = \frac{1}{2}(1+3w)-\frac{1}{2}(3+3w)\; \Omega_\Lambda
\ee
In particular, 
\ba\label{2.11}
&&q= 1-2\; \Omega_\Lambda  \; \; \; \; \; \; \; \; \; \; \;
\textrm{ for } \; \;  w=1/3\\
&&q= (1-3 \; \Omega_\Lambda)/2 \; \; \; \; \;  \textrm{ for }\; \;  w=0 
\ea
for a radiation and a matter dominated Universe, respectively.
\subsection{The cutoff identification}
Up to now $G$ and $\Lambda$ were prescribed functions of time. Now we induce their
$t$-dependence by means of the cutoff identification
\be\label{2.12}
k(t) =\xi H (t)
\ee
from a given RG trajectory of the Einstein-Hilbert truncation, 
$\Big (g(k),\lambda(k) \Big )$. Here $\xi$ is a fixed positive constant of order unity.
Eq.(\ref{2.12}) is a natural choice since in a Robertson-Walker geometry
the Hubble parameter measures the curvature of spacetime; its inverse $H^{-1}$ defines 
the size of the ``Einstein elevator''. With $G(k)=g(k)/k^2$ and
$\Lambda(k) = \lambda(k) k^2$   we have 
\ba\label{2.13}
&&G(t) = \frac{g(\xi H(t))}{\xi^2 \; H(t)^2}\\[2mm]
&&\Lambda(t) = \xi^2 \; H(t)^2 \; \lambda(\xi H(t))\nonumber
\ea
The algorithm for solving the cosmological evolution equation assumes 
the following form now:

{\it Let $ \Big ( g(k),\lambda(k) \Big )$ be a prescribed RG trajectory 
and $H(t)$ a solution of }
\be\label{2.14}
\dot{H}(t)= -\frac{1}{2}(3+3w)H(t)^2\Big [ 1-\frac{1}{3} \; \xi^2 \; \lambda(\xi H(t)) \Big]
\ee
{\it Let $\rho(t)$ be defined in terms of this solution by }
\be\label{2.155}
\rho(t) = \frac{3 \; \xi^2}{8\pi\; g(\xi H(t))} 
\Big [ 1-\frac{1}{3}\; \xi^2 \;  \lambda(\xi H(t)) \Big ] \, H(t)^4
\ee
{\it Then the pair $\Big( H(t),\rho(t) \Big )$ 
is a solution of the system} (\ref{2.5a}), (\ref{2.5b})
{\it for the time dependence of $G$ and $\Lambda$ 
given by} (\ref{2.13}) {\it and the equation of state $p=w\rho$, provided $H(t)\not =0$. }

Later on we shall apply this algorithm to the various cosmological epochs of interest.
Sometimes it is more convenient to regard $H$ and $\rho$ functions of the scale factor 
rather than time. Exploiting that $a(dH/da)=(dH/dt)/H$ we see that Eq.(\ref{2.14})
implies the following somewhat simpler differential equation for $H=H(a)$:
\be\label{2.140}
a\frac{dH(a)}{da}=-\frac{1}{2}(3+3w) \, H(a) \, \Big [ 1-\frac{1}{3}\, \xi^2\, \lambda(\xi H(a))\Big ]
\ee
Or, using logarithmic variables,
\be\label{2.141}
\frac{d \ln H}{d \ln a}=-\frac{1}{2}(3+3w)\Big [ 1-\frac{1}{3}\, \xi^2\, \lambda(\xi H)\Big ]
\ee
The latter equation is particularly convenient for the numerics.
\subsection{Cosmology on theory space}
The theory space of the Einstein-Hilbert truncation can be identified with  a part
of the $g$-$\lambda$--plane. There are certain quantities of cosmological interest whose values
at time $t$ depend only  on the point of theory space the Universe passes at this time, but
 not on $t$ or on the dynamics directly.
Such quantities are functions of  $g$ and $\lambda$. Examples are $\OL$, $\OM$, and $q$ which
actually are functions of $\lambda$ alone:
\ba\label{2.15}
&&\OL(\lambda) = \frac{\xi^2}{3}\; \lambda\\[2mm]
&&\OM(\lambda) = 1-\frac{\xi^2}{3}\; \lambda\nonumber\\[2mm]
&&q(\lambda) = \unm \Big [ (1+3w)-(1+w)\; \xi^2\; \lambda \Big ]\nonumber
\ea
Remarkably, whether or not the Universe decelerates at some time $t$ depends only on the value
of $\lambda$  at this time. Defining  the ``zero acceleration'' value
\be\label{2.16}
\laz(w) = \frac{1+3w}{1+w}\, \frac{1}{\xi^2}
\ee
we have deceleration $(q>0)$ if $\lambda<\laz$ and acceleration $(q<0)$ if 
$\lambda > \laz$. For $\lambda=\laz$ we get $q=0$ and $a\propto t$ therefore.
The ``zero acceleration  line'' $\{ (g, \laz) |-\infty<g<\infty\}$ divides the
$g$-$\lambda$--plane in two parts, with deceleration on its left and acceleration on its right.

Another line on the $g$-$\lambda$--plane which is of special significance is the so-called
``$\Omega$-line`` [II] along which, by definition, $\rho=0$, 
\ie $\OM=0$. By $(\ref{2.15})$, 
this line  $\{ (g, \lambda_{\Omega}) |-\infty<g<\infty\}$ is parallel to the zero acceleration line, 
with 
\be\label{2.17}
\lambda_\Omega = \frac{3}{\xi^2} \;\;\;\;\;\;\; \; \; \textrm{for all} \;\; w.
\ee
Provided $\lambda_\Omega < 1/2$, cosmologies with an eternal expansion 
and a corresponding dilution of their matter 
contents terminate on the $\Omega$-line for $t\rightarrow \infty$. If $\lambda_\Omega> 1/2$ this
line is of no physical importance since it lies in a region where the Einstein-Hilbert truncation is not
valid.  (For the corresponding discussion when the consistency condition 
is imposed we refer to [II]. )
\section{Entropy generation}
\renewcommand{\theequation}{3.\arabic{equation}}
\setcounter{equation}{0}
Let us return to the modified continuity equation (\ref{2.5b}). After multiplication by $a^3$
it reads
\be\label{3.1}
[\dot\rho + 3H(\rho +p)] \; a^3 = \calp(t)
\ee
where we defined
\be\label{3.2}
\calp\equiv -\Big ( \frac{\dot{\Lambda}+8 \pi\; \rho \; \dot{G}}{8\pi \; G} \Big ) a^3
\ee
Without assuming any particular equation of state Eq.(\ref{3.1}) can be
rewritten as 
\be\label{3.3}
\frac{d}{dt} (\rho a^3) +p\frac{d}{dt}(a^3) = \calp(t)
\ee
The interpretation of this equation is as follows. 
Let us consider a unit {\it coordinate}, 
i.e. comoving volume in the Robertson-Walker spacetime. Its corresponding  
{\it proper} volume is $V=a^3$
and its energy contents is $U=\rho a^3$. The rate of change of these quantities is subject to 
(\ref{3.3}): 
\be\label{3.4}
\frac{dU}{dt}+p\frac{dV}{dt}=\calp(t)
\ee
In classical cosmology where $\calp\equiv 0$ this equation together with the standard thermodynamic
relation $dU+pdV=TdS$ is used to conclude that the expansion of the Universe is adiabatic, \ie
the entropy inside a comoving volume does not change as the Universe expands, $dS/dt=0$.

Here and in the following we write $S\equiv s \, a^3$ for the entropy carried by the matter inside
a unit comoving volume and $s$ for the corresponding proper entropy density.

When $\Lambda$ and $G$ are time dependent, $\calp$ is nonzero and we interpret (\ref{3.4})
as describing the process of energy (or ``heat'') exchange between the scalar fields $\Lambda$ 
and $G$  and the ordinary matter. This interaction causes $S$ to change:
\be\label{3.5}
T\frac{dS}{dt}=T\frac{d}{dt}(s a^3)=\calp(t)
\ee
The actual rate of change of the comoving entropy is 
\be\label{3.6}
\frac{dS}{dt}=\frac{d}{dt}(s a^3)= \clp (t)
\ee
where 
\be\label{3.7}
\clp \equiv \calp /T
\ee 
If $T$ is known as a function of $t$ we can integrate (\ref{3.5}) to obtain $S=S(t)$. 
In the RG improved cosmologies the entropy production rate per comoving 
volume 
\be\label{3.7}
\clp(t) = - \Big [ \frac{ \dot\Lambda+8\pi \; \rho \; \dot G}{8\pi\;  G} \Big ] \frac{a^3}{T}
\ee
is nonzero because the gravitational ``constants''  $\Lambda$ and $G$ have acquired a time 
dependence. 

For a given solution to the coupled system of RG and cosmological 
equations it is sometimes more convenient to calculate $\clp (t)$ from the LHS of the 
modified continuity equation rather than its RHS (\ref{3.7}): 
\be\label{3.8}
[\dot\rho + 3H(\rho +p)] \frac{a^3}{T} = \clp(t)
\ee

If $S$ is to increase with time, by (\ref{3.7}), we need that $\dot\Lambda +8\pi \dot{G}<0$.
During most epochs of the RG improved cosmologies we have $\dot\Lambda \leq 0$ and
$\dot G\geq 0$. The decreasing $\Lambda$ and the increasing $G$ have antagonistic effects 
therefore. We shall see that in the physically realistic cases $\Lambda$ predominates so that
there is indeed a transfer of energy from the vacuum to the matter sector rather than
vice versa.

Clearly we can convert the heat exchanged, $TdS$, to an entropy change only if the dependence
of the temperature $T$ on the other thermodynamical quantities, in particular $\rho$ and $p$
is known.  For this reason we shall now make the following assumption about the matter system and its
(non-equilibrium!)  dynamics:

{\it The matter system is assumed to consist 
of $n_{\rm eff}$ species of effectively massless degrees of freedom
which all have the same temperature $T$. The equation of state is $p=\rho/3$, 
\ie $w=1/3$, and $\rho$ depends on $T$ as 
\be\label{3.9}
\rho(T) =\kappa^4 \; T^4 , \;\;\;\;\;\;\; \kappa\equiv (\pi^2 \; n_{\rm eff}/30)^{1/4}
\ee
No assumption is made about the relation $s=s(T)$.}

The first assumption, radiation dominance and equal temperature, is plausible since we shall find
that there is no significant entropy production any more once $H(t)$ has dropped substantially below
$\mp$, after the crossover from the NGFP to the GFP.

The second assumption, eq.(\ref{3.9}), amounts to the hypothesis formulated in the introduction. 
While entropy generation is a non-adiabatic process we assume, following Lima \cite{lima1},
that the non-adiabaticity is as small as possible. More precisely, the approximation is that the 
{\it equilibrium} relations among $\rho$, $p$, and $T$ are still valid in the
non-equilibrium situation of a cosmology with entropy production. In this sense, (\ref{3.9})
is the extrapolation of the standard relation (\ref{1.3a}) to a ``slightly non-adiabatic''
process.

Note that while we used (\ref{1.3c}) in relating $\clp(t)$ to the entropy production and also
postulated eq.(\ref{1.3a}), we do not assume the validity of the formula for
the entropy density, eq.(\ref{1.3b}), a priori. We shall see that the latter is an automatic
consequence of the cosmological equations. 

To make the picture as clear as possible we shall
neglect in the following all ordinary dissipative processes in the cosmological fluid.

Using $p=\rho/3$ and (\ref{3.9}) in (\ref{3.8}) the entropy production rate can be evaluated as follows:
\ba\label{3.20}
&\clp(t)=& \kappa\;  \Big [ a^3 \rho^{-1/4} \; \dot \rho + 4\; a^3 \; H \; \rho^{3/4} \Big ]\\[2mm]
& \; \; \; \; \; \;\;\;=& \frac{4}{3} \; \kappa \; \Big [ a^3\; \frac{d}{dt}(\rho^{3/4}) 
+ 3 \; \dot{a} \; a^2 \; \rho^{3/4} \Big ]\nonumber\\[2mm]
& \; \; \; \; \; \;\;\;=& \; \frac{4}{3} \; \kappa \Big [ \; a^3 \; \frac{d}{dt}(\rho^{3/4}) 
+ \rho^{3/4}  \frac{d}{dt}(a^3) \Big ]\nonumber
\ea
Remarkably, $\clp$ turns out to be a total time derivative:
\be\label{3.21}
\clp(t) = \frac{d}{dt} \; \Big [ \frac{4}{3} \; \kappa \; a^3 \; \rho^{3/4} \Big ]
\ee
Therefore we can immediately integrate (\ref{3.5}) and obtain 
\be\label{3.22}
S(t)=\frac{4}{3}\; \kappa \; a^3\; \rho^{3/4} +S_{\rm c}
\ee
or, in terms of the proper entropy density, 
\be\label{3.23}
s(t) = \frac{4}{3} \; \kappa\; \rho(t)^{3/4} +\frac{S_{\rm c}}{a(t)^3}
\ee
Here $\ssc$ is a constant of integration. In terms of $T$, using 
(\ref{3.9}) again, 
\be\label{3.24}
s(t) = \frac{2\pi^2 }{45} \; n_{\rm eff} \; T(t)^3 +\frac{S_{\rm c}}{a(t)^3}
\ee

The final result (\ref{3.24}) is very remarkable for at least two reasons. First, 
for $\ssc=0$, Eq.(\ref{3.24}) has exactly the form (\ref{1.3b}) which is valid for radiation
in equilibrium. Note that we did not postulate this relationship, only the $\rho(T)$--law
was assumed. The equilibrium formula $s\propto T^3$ was {\it derived} from the 
cosmological equations, \ie the modified conservation law. This result makes the hypothesis
``non-adiabatic, but as little as possible'' selfconsistent. 

Second, if $\lim_{t\rightarrow 0} \; a(t) \rho(t)^{1/4}=0$, which is  actually the
case for the most interesting class of cosmologies we shall find, then $S(t\rightarrow 0)=S_c$
by eq.(\ref{3.22}). As we mentioned in the introduction, the most plausible initial value
of $S$ is $S=0$ which means a vanishing constant of integration $S_c$ here. But then,
with $S_c=0$, (\ref{3.22}) tells us that the {\it entire} entropy carried by the 
massless degrees of freedom is due to the RG running. So it indeed seems to 
be true that the entropy of the CMBR photons we observe today is due to a coarse graining but,
unexpectedly, not a coarse graining of the matter degrees of freedom but rather of the
gravitational ones which determine the background spacetime the photons propagate on.

We close this section with various comments.
As for the interpretation of the function $\clp(t)$, let us remark that it also measures
the deviations from the classical laws $a^4\rho = const$ and $aT=const$, respectively,  since
we have $\clp = \frac{4}{3}\; \kappa \; d (a^4 \rho)^{3/4} /dt= \frac{4}{3}\kappa^4\;  {d}(aT)^3 /dt$.

Both in classical and in improved cosmology with the ``consistency condition'' imposed
the quantity $\clm \equiv 8\pi a^4\rho$ is conserved in time \cite{cosmo1}. If energy transfer
is permitted and the entropy of the ordinary matter grows, $\clm$ increases as well. This is obvious
from
\be\label{3.30}
\frac{d}{dt} \clm (t)^{3/4} = \frac{3}{4\kappa}(8\pi)^{3/4}\; \clp (t)
\ee
or, in integrated form, $S(t)=\frac{4}{3}\kappa (8\pi)^{-3/4} {\cal M}(t)^{3/4} +\ssc$ .

In a spatially flat Robertson-Walker spacetime the overall scale of $a(t)$ has no physical
significance. If $\clm$ is time independent, we can fix this gauge ambiguity by picking a 
specific value of $\clm$ and expressing $a(t)$ correspondingly. For instance, parametrized in this
way, the scale factor of the classical FRW cosmology with $\Lambda=0$, $w=1/3$ reads
\cite{cosmo1}
\be\label{3.31}
a(t)=  [ 4\barg \clm /3 ]^{1/4} \; \sqrt{t}
\ee
If, during the expansion, $\clm$ increases slowly, eq.(\ref{3.31}) tells us that the 
expansion is actually {\it faster} than estimated classically. Of course, what we actually 
have to do in order to  find the corrected $a(t)$ is to solve the improved field equations
and not insert $\clm=\clm(t)$ into the classical solution, in particular when the 
change of $\clm$ is not ``slow''. Nevertheless, this simple argument makes it clear that 
entropy production implies an increase of $\clm$ which in turns implies an extra increase
of the scale factor. This latter increase, or ``inflation'', is a pure quantum effect.
The explicit solutions to which we turn next will confirm this picture.
\section{Explicit analytical solutions}
\renewcommand{\theequation}{4.\arabic{equation}}
\setcounter{equation}{0}
\setcounter{figure}{1}
In this and the next section we explicitly solve the cosmological evolution equations pertaining to
type IIIa trajectories. In this section we discuss analytical solution for the fixed
point, the $k^4$-, and the classical regime in turn, and then  obtain complete
cosmological histories by numerical methods in the next section. 
\subsection{The fixed point regime}
For ``$k=\infty$'' every trajectory stays very close to the NGFP with constant values
$g(k)\equiv \lst$ and $\lambda(k)\equiv \lst$. In this regime the differential equation
(\ref{2.14}) reads 
\be\label{4.9}
\dot{H}(t) = -\alpha^{-1} \; H^2(t)
\ee
with the constant
\be\label{4.10}
\alpha \equiv \frac{2}{(3+3w)[1-\lst \xi^2/3]}
\ee
Eq.(\ref{4.9}) describes a cosmology with an initial singularity. Fixing the constant of
integration such that this singularity occurs at $t=0$, the unique solution to (\ref{4.9})
reads 
\be\label{4.11}
H(t) = \frac{\alpha}{t}
\ee
which integrates to $a(t)\propto t^\alpha$. The exponent $\alpha$ depends on the combination
$\lst \xi^2$; by eq.(\ref{2.15}) this is essentially the $\OL$-value in the NGFP regime:
\be\label{4.110}
\OLS =\lst \; {\xi^2}/{3}
\ee
Henceforth we shall always eliminate $\xi$ in favor of the more physical parameter $\OLS$.
Then, 
\be\label{4.12}
\alpha = \frac{2}{(3+3w)(1-\OLS)}
\ee
Using $\OLS$ as the free parameter  which distinguishes different solutions, the fixed point
cosmologies are characterized by the following power laws:
\begin{subequations}\label{4.13}
\ba\label{4.13a}
&& a(t) = A \; t^\alpha \; \; , \;\;\;\;\; A>0 \\[2mm]\label{4.13b}
&&\rho(t) = \frac{\hat{\rho}}{t^4}\; ,  \;\;\;\;\;
\hat{\rho}=\frac{2 \; \OLS}{ 9 \pi \; \gst\lst \; (1+w)^4 \; (1-\OLS)^3}\\[2mm]\label{4.13c}
&&G(t)=\frac{3 \; \gst\lst \; (1+w)^2 \; (1-\OLS)^2}{4 \; \OLS} \; t^2\\ [2mm]\label{4.13d}
&&\Lambda(t) = \frac{4 \; \OLS}{3 \; (1+w)^2\; (1-\OLS)^2} \; \frac{1}{t^2}
\ea
\end{subequations}
Eq.(\ref{4.13b}) follows from  (\ref{2.15}) by inserting (\ref{4.11}), while
(\ref{4.13c}) and (\ref{4.13d}) are eqs.(\ref{2.13}) for $H=\alpha/t$ and constant
values of $g$ and $\lambda$. The parameter $\xi$ has been eliminated in favor of $\OLS$
everywhere. Note that the RG data enter the solution (\ref{4.13}) only via the universal
\cite{oliver1,frank1} product $\gst\lst$. 

The solution (\ref{4.13}) has time independent values
of $\OL=\OLS$, $\OM=1-\OLS$, and 
\be\label{4.14}
q=\frac{1}{\alpha}-1=\unm [1+3w-(3+3w)\; \OLS]
\ee
Eliminating $\xi$ from (\ref{2.16}) and (\ref{2.17}) we can express $\laz$ and
$\lambda_\Omega$ in terms of $\OLS$:
\ba\label{4.15}
&&\laz (w)=\frac{1+3w}{3+3w} \, \frac{\lst}{\OLS}\\[2mm]
&& \lambda_\Omega= \frac{\lst}{\OLS}
\ea
In the radiation dominated case we have
\be\label{4.17}
\laz(1/3)=\frac{\lst}{2\; \OLS}< \lambda_\Omega = \frac{\lst}{\OLS}
\ee
so that for any value of $\OLS \in (0,1)$ the zero acceleration line is on the left of the
$\Omega$-line. The relative location of the NGFP depends on whether $\OLS$ is bigger or smaller
than $1/2$:
\ba\label{4.18}
&&\laz(1/3)\leq\lst<\lambda_\Omega \;\;\;\;\;\; \textrm{if} \;\;\; \OLS\in [1/2, 1)\\[2mm]
&&\lst<\laz(1/3)<\lambda_\Omega \;\;\;\;\;\; \textrm{if}\;\;\;  \OLS\in (0,1/2)\nonumber
\ea
\begin{figure}[t]\label{figure:fig2}
\begin{center}
\includegraphics[width=0.85\textwidth]{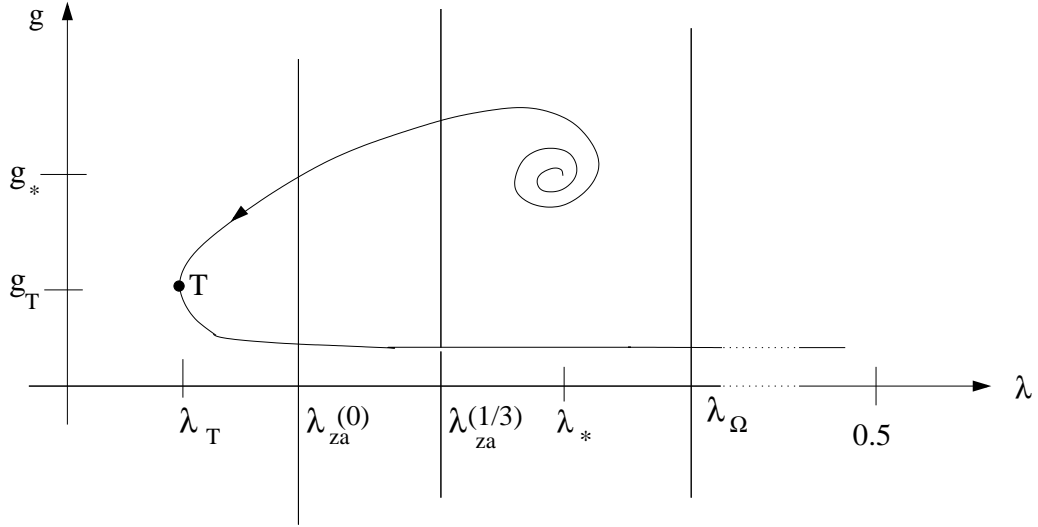}
\end{center}
\caption{ A trajectory of the type IIIa. The vertical lines are the ``$\Omega$-line''
and the zero-acceleration lines for $w=0$ and $w=1/3$. The $\Omega$-line is not necessarily in the 
physical part of parameter space $(\lambda < 1/2)$.} 
\label{fig2}
\end{figure}
The $\Omega$-line is relevant only if it is in the physical part of parameter space
$(\lambda < 1/2)$. If indeed $\lambda_\Omega < 1/2$, the first case of (\ref{4.18}) 
corresponds to the sketch in Fig.~2. 

When $\laz$ is smaller than
$\lst$, the fixed point and the spiraling regime of the RG trajectory close to the NGFP
correspond to an accelerating epoch of the Universe. (In the second case it would be decelerating.)
If $1/2 < \OLS < 1$, the NGFP regime is an epoch of ``power law-inflation'' $a\propto t^\alpha$, 
with $\alpha >1$ and $q<0$ :
\ba\label{4.180}
&& \alpha(w=1/3) = (2-2 \; \OLS)^{-1}\\[2mm]
&& q(w=1/3) = 1-2 \; \OLS
\ea 
Note that for $\OLS \nearrow 1$ the exponent $\alpha(1/3)$ becomes very large and $q(1/3)$
approaches $-1$.

The eqs.(\ref{4.13}) describe a one-parameter family of cosmologies labeled by $\OLS$. The
solution exists (and has $\alpha >0$, $\rho>0$) for any value of $\OLS$ in the interval $(0,1)$.
The possibility of freely\footnote{However, physically plausible values of $\OLS$ should be such that
$\xi^2=3\OLS/\lst$ does not assume unnaturally small or large values.} choosing $\OLS$ is the
main new feature as compared to [I] where the 
``consistency condition'' (\ref{2.6b}) had been imposed.
The consequence of imposing (\ref{2.6b}) is exactly to eliminate all solutions of the family
(\ref{4.13}) except the one for $\OLS=1/2$. In fact, it is easily checked that (\ref{4.13})
for the special case $\OLS=1/2 $ is identical to the fixed point solution found in [I].
The solutions for $\OLS\not=1/2$ are new.

The case $\OLS=1/2$ is special for a variety of reasons. It has, for instance, 
equal relative matter and  vacuum energy densities, $\OMS=\OLS=1/2$, 
vanishing deceleration parameter $(q=0)$ corresponding to a linear expansion $a\propto t$, 
$\alpha =1$, and the NGFP sits precisely on the zero acceleration line in this case [II].
If the ``consistency condition'' is imposed, ${\cal M}\equiv 8\pi \rho a^{3+3w}$ is time
independent, while in the present more general framework it is not. For the above fixed point
solutions one finds ${\cal M}(t) \propto t^{2/(1-\OLS) -4}$ which is constant in the 
exceptional case $\OLS=1/2$ only. In [I] we reexpressed the constant $A$ of (\ref{4.13})
in terms of ${\cal M}$ which is no longer possible here. 

If the NGFP expansion $a(t)\propto t^\alpha$ is realized for $t\rightarrow 0$, the Universe
has no particle horizon if $\alpha \geq 1$, but does have a horizon of radius $d_{H}=t/(1-\alpha)$
if $\alpha < 1$. In the case of $w=1/3$ this means that there is  a horizon for 
$\OLS< 1/2$, but none if $\OLS \geq 1/2$.

If $w=1/3$, the discussion of Sec.3 on entropy generation applies to the NGFP regime
of the improved cosmology. The corresponding rate of entropy production reads
\be\label{4.20}
\clp(t) = 4 \kappa \; (\alpha-1) \; A^3\; \widehat{\rho}^{3/4} \; t^{3\alpha -4}
\ee
where $\alpha \equiv \alpha (w=1/3) $ is given by (\ref{4.180}), and
\be\label{4.21}
\widehat{\rho}\equiv \widehat{\rho}\; (w=1/3) =
\frac{9 \; \OLS}{128  \pi \; \gst \lst \; (1-\OLS)^3}
\ee
As expected, $\clp$ vanishes identically if $\alpha=1$, \ie $\OLS=1/2$. In this case the 
solution obeys the ``consistency condition'' and no energy is exchanged
with the matter system. For the entropy per unit comoving volume we
find, if $\alpha \not =1$, 
\be\label{4.22}
S(t) = \ssc + \frac{4}{3} \kappa \; A^3 \; \widehat{\rho}^{3/4}\; t^{3(\alpha -1)}
\ee
and the corresponding proper entropy density is 
\be\label{4.23}
s(t) =\frac{\ssc}{A^3 \; t^{3\alpha}}+ \frac{4\kappa \; \widehat\rho ^{3/4}}{3\;  t^3}
\ee
Here $\ssc$ is an undetermined constant of integration. 
The temperature behaves as $1/t$ for any value of $\alpha$:
\be\label{4.230}
T(t) = \frac{\widehat{\rho}^{1/4}}{\kappa \; t}
\ee

For the discussion of the entropy we must distinguish 3 qualitatively different cases.
They differ in particular with respect to the sign of $\clp (t)$, the behavior of $S(t)$
close to the initial singularity,  and with
respect to the relative importance of the running cosmological and Newton constant. 
(Cf. the remark after eq.(\ref{3.8}) .)

\noindent
{\bf (a) The case $\alpha >1$, \ie $1/2< \OLS < 1$:} 

\noindent 
Here $\clp(t)>0$ so that the 
entropy and energy content of the matter system increases with time. By eq.(\ref{3.7}), 
$\clp >0$ implies $\dot\Lambda +8\pi  \rho  \dot G <0$. Since $\dot\Lambda <0$ but $\dot G>0$
in the NGFP regime, the energy exchange is predominantly due to the decrease of $\Lambda$
while the increase of $G$ is subdominant in this respect.

The comoving entropy (\ref{4.22}) has a finite limit for $t\rightarrow 0$, 
$S(t\rightarrow 0) =\ssc$, and $S(t)$ grows monotonically for $t>0$. If $\ssc=0$,  
which would be the most
natural value in view of the discussion in the introduction, {\it all}  of the entropy carried
by the matter fields is due to the energy injection from $\Lambda$. But even if $\ssc \not = 0$, 
any such nonzero initial value will be irrelevant at a sufficiently late time (at least if this time
is smaller than $\approx \tp$ above which (\ref{4.22}) is invalid) .

\noindent
{\bf (b) The case $\alpha < 1$, \ie $0<\OLS< 1/2$:}

\noindent
Here $\clp(t)<0$ so that the energy and entropy of matter decreases. Since $\clp <0$ 
amounts to $\dot\Lambda +8\pi  \rho  \dot G>0$, the dominant physical effect is the increase of 
$G$ with time, the counteracting decrease of $\Lambda$ is less important.
The comoving entropy starts out from an infinitely positive value at the initial 
singularity, $S(t\rightarrow 0) \rightarrow +\infty$, and decreases thereafter.

\noindent
{\bf (c) The case $ \alpha=1$, \ie $\OLS=1/2$:}

\noindent
Here $\clp(t)\equiv 0$, $S(t)=const$, and $\dot\Lambda +8\pi  \rho  \dot G=0$. 
The effect of a decreasing
$\Lambda$ and increasing $G$ cancel exactly so that there is not net energy exchange. 
\subsection{The $k^4$-regime}
The ``$k^4$-regime'' of a type IIIa trajectory is its part which can be described 
by the linearization about the GFP where $\Lambda$ has a quartic $k$-dependence.
The corresponding trajectory $\Big (\lambda(k), g(k) \Big )$ is given in eq.(\ref{1.11}).
We shall set $\Lambda_0=0$ there which is a good approximation as long as $k^2\gg \Lambda_0$ .
Since we are mostly interested in the entropy production in the early Universe 
(soon after the Planck regime) this approximation is sufficient for our purposes.
(For the separatrix which has $\Lambda_0=0$ it represents no approximation at all.)

In this regime we have $G=const$ approximately and 
\be\label{4.30}
\Lambda = \nu \; \xi^4 \; \bar{G}\; H^4
\ee
As $\Lambda = 3  \OL  H^2$, this implies directly
\ba\label{4.31}
&& \OL = L^2\; H^2, \;\;\;\;\;\; \OM = 1-L^2 \; H^2 \\[2mm]
&& q= \unm \Big [ 1+3w-(3+3w) L^2 H^2 \Big ] \label{4.32}
\ea
where the scale is set by the  quantity
\be\label{4.33}
L\equiv \sqrt{\nu/3} \; \xi^2 \sqrt{\barg}
\ee
which is a length of the order of the Planck length.  The differential equation (\ref{2.14})
for $H(t)$ reads in the present case
\be\label{4.34}
\dot{H}=-\alpha_0^{-1} \; H^2 \; (1-L^2 H^2)
\ee
where 
\be\label{4.35}
\alpha_0\equiv \frac{2}{3+3w}
\ee
The general solution can be found in the form $t=t(H)$ by a simple  integration, but the
inversion is not elementary:
\be\label{4.36}
t(H) = \alpha_0 \; [ H^{-1}- L\; \textrm{artanh}(LH) ] +const
\ee
It is easier to solve the equation (\ref{2.140}) for $H=H(a)$ which reads in the case at hand
\be\label{4.37}
a\frac{dH}{da}  = -\alpha_0^{-1} \; H (1-L^2 H^2)
\ee
Its general solution is given by 
\be\label{4.38}
H(a) = \frac{1}{L} \Big [ 1+ \Big (\frac{a}{\tilde a} \Big )^{3+3w} \Big ]^{-1/2}
\ee
where $\tilde a$ is a constant. As an aside we note that the cosmology described
by (\ref{4.38}), when taken seriously for any $a$, does not have an initial singularity
since the RHS of (\ref{4.38}) is bounded above. This observation is of no relevance in
the present context, however, because once $H$ gets larger than the Planck mass 
(\ref{4.38}) becomes invalid and the fixed point solution takes over, which does
have a ``big bang''. 

Inserting (\ref{4.38}) into (\ref{pippo}) we obtain the energy density as a function of 
the scale factor:
\be\label{4.39}
\rho (a) = \frac{3}{8\pi \barg L^2} \; \Big ( \frac{a}{\tilde a} \Big )^{3+3w}\;
\Big [ 1+\Big (\frac{a}{\tilde a} \Big )^{3+3w}\Big ] ^{-2}
\ee
In a realistic cosmology the epoch during which (\ref{4.39}) is valid is radiation dominated so 
that we should set $w=1/3$. Since $L=O(\lp)$ we see that $\tilde a$ is the scale factor
at which $k$,$H=O(\mp)$ and $\rho=O(\mp^4)$. Eq.(\ref{4.39}) implies the following $a$-dependence
of the entropy:
\be\label{4.40}
S(a) = S(\tilde a)+
\frac{4}{3}\kappa \tilde{a}^3 \rho(\tilde{a})^{3/4} \; 
\bigg \{ 2\sqrt{2} \Big (\frac{a}{\tilde a}\Big )^6 \Big [ 1+ 
\Big (\frac{a}{\tilde a}\Big )^4 \Big ]^{-3/2}-1 \bigg \}
\ee
This formula predicts a significant entropy production only near $a\approx \tilde a$, 
$k \approx \mp$, \ie in the crossover regime. There the function inside the
curly brackets of eq.(\ref{4.40}) grows from $-1$ at $a\ll \tilde a$ to 
$+2\sqrt{2}$ at $a\gg \tilde a$.
\subsection{Classical regime and the value of $\tilde a$}
For $k\ll k_{\rm T}$, in the ``IR-regime'', Nature's RG trajectory has a long classical regime
where $G\approx const$ and $\Lambda \approx const$. In this regime the functions $H(t)$ and
$\rho(t)$ are well known, of course\footnote{See for instance Appendix B of [I] where the 
present notation is employed.}. Here we quote only the result for a negligible $\Lambda$:
\be\label{4.45}
a(t) = \Big [ \frac{3}{4} (1+w)^2{\cal M} \barg \Big ]^{1/(3+3w)} \; t^{2/(3+3w)}
\ee
There is no entropy production in the classical regime, $\clp=0$. The scale
factor (\ref{4.45}) corresponds to 
\be\label{4.46}
H(a) = \sqrt{\frac{1}{3}{\cal M}\barg}\; a^{-(3+3w)/2} 
\ee
with the invariant ${\cal M}\equiv 8\pi \rho a^{3+3w}$; its value  can be determined by astrophysical
observations in the late Universe.

We can use (\ref{4.46}) in order to fix the constant $\tilde a$ of eq.(\ref{4.38}) valid in 
the preceding ``$k^4$-regime''. For $a\gg \tilde a$ the latter equation yields 
\be\label{4.47}
H(a) \approx \frac{1}{L} \; \Big ( \frac{a}{\tilde a} \Big )^{-(3+3w)/2}
\ee
Identifying (\ref{4.46}) and (\ref{4.47}) we read off that $\tilde a$ can be expressed
in terms of ${\cal M}$ as 
\be\label{4.48}
\tilde a =  [ \nu \; \xi^4 \; \barg \; {\cal M}/9 ]^{1/(3+3w)}
\ee
Since both the $k^4$- and the early classical regime are in the radiation dominated
epoch the above comparison requires $w=1/3$.

Matching the observed data against the classical FRW cosmology in the usual way one finds that the 
quantity ${\cal M}$ for the radiation dominated era 
of the Universe we live in is approximately given by 
\be\label{4.49}
{\cal M}\approx (10^{-30} a_0 /\lp)^4
\ee
with $\lp\equiv \mp^{-1} \equiv \sqrt{\barg}$, and $a_0$ the scale factor today.
Since $\nu$, $\xi=O(1)$, the order of magnitude estimate (\ref{4.49}) implies that
\be\label{4.50}
{\tilde a}\approx 10^{-30} \; a_0
\ee
Hence in the $k^4$-regime, approximately,
\be\label{4.51}
H(a)\approx \mp \; \Big [1+c\; (10^{30}a/a_0)^4 \Big ]^{-1/2}
\ee
where $c$ is a constant of order unity. 

Note that using (\ref{4.49}) in (\ref{4.45}) for $w=1/3$ we get that, classically, 
and up to factors of order unity, 
\be\label{4.52}
a(t) \approx 10^{-30} a_0 \sqrt{t/\tp}
\ee
Therefore $\tilde a $ has the interpretation of the scale factor $a(\tp)$  predicted by classical cosmology 
for the time $t=\tp$ . It is just at this scale, however, where according to (\ref{4.51})
deviations from the classical behavior start to occur.
\section{Complete cosmological histories}
\renewcommand{\theequation}{5.\arabic{equation}}
\setcounter{equation}{0}
In this section we construct complete cosmologies by numerically integrating the coupled
system of RG and cosmological evolution equations from the ``big bang'' up to asymptotically late
times. As for  the RG equations we use the same cutoff scheme, the sharp cutoff, as in [II]
in order to facilitate the comparison. In 4 dimensions the flow equations for the Einstein-Hilbert
truncation with a sharp cutoff read \cite{frank1} 
\ba\label{5.1}
&&k\partial_k g(k) = \beta_g (g,\lambda) \equiv [ 2+\ett(g,\lambda)]\; g\\[2mm] 
&&k\partial_k \lambda(k) = \beta_\lambda(g,\lambda)\nonumber
\ea
with $\beta_\lambda$ and $\ett$ given by 
\ba\label{5.2}
&&\beta_\lambda(g,\lambda) = -(2-\ett)\lambda -\frac{g}{\pi}\Big [ 5\, \ln(1-2\lambda)
-\varphi_2 + \frac{5}{4}\, \ett \Big ]\\[2mm]
&&\ett(g,\lambda) = - \frac{2 \, g}{6\pi +5\, g} 
\Big [ \frac{18}{1-2\lambda}+5\, \ln(1-2\lambda)-\varphi_1 +6 \Big ] . \label{5.3}
\ea
As in  [II] we use the variant of the sharp cutoff with ``shape parameter'' $s=1$ for which $\varphi_1=\zeta(2)$, 
$\varphi_2 =2 \zeta(3)$, see \cite{frank1}. The NGFP is at $\gst=0.403$, $\lst=0.330$ then.

In particular in numerical computations employing Nature's RG trajectory which comprises
very many orders of magnitude it is advantageous to use logarithmic variables. We normalize
them with respect to their value at the turning point and express the RG time, the scale
factor, the cosmological time, and the Hubble parameter by, respectively, 
\ba\nonumber
&&\tau \equiv \ln (k/\kat)\\[2mm]
&& x \equiv \ln (a/\aT)\nonumber\\[2mm]
&& y \equiv \ln (t/\tT)\label{5.4}\\[2mm]
&& {\cal U} \equiv \ln (H/\hT)\nonumber
\ea
By definition, $x$ and $y$ are negative in what we call the ``UV-regime'', the upper
branch of the trajectory where $k>\kat$, and positive in the ``IR-regime'', the
lower branch with $k<\kat$. For $\tau$ and ${\cal U}$ it is the other way around.
The variable $x$ is the number of ``e-folds'' relative to the size of the Universe
when it passes the turning point of the underlying RG trajectory. In these variables,
the cutoff identification $k=\xi H$ implies ${\cal U}=\tau$. 

In a regime with power law expansion $a\propto t^\alpha$ we have 
\be\label{5.5}
x=\alpha \; y , \;\;\;\;\;\; \UU=\tau = -y
\ee
For instance, in the NGFP regime with $w=1/3$, 
\be\label{5.6}
x=(2-2 \; \OLS)^{-1} \; y, \;\;\;\;\;\; \UU = \tau =-y = -(2-2\OLS)\; x
\ee
while for a classical FRW cosmology with $\Lambda=0$, 
\ba\label{5.7}
&&x= \unm \; y, \;\;\;\;\;\ \UU=\tau =-y=-2\; x \; \;\;  \text{if}\; \; \; w=1/3\\[2mm]
&&x= \frac{2}{3}\; y, \;\;\;\;\;\; \UU=\tau =-y=-\frac{3}{2}\; x \;\;\; \text{if}\;\;\; w=0
\ea
We shall need these simple relations repeatedly.

For reasons of numerical stability it is advantageous to integrate the coupled
system of the RG and cosmological differential  equations not with respect to $t$ 
but rather  $x$. Let us write 
\be\label{5.9}
\gs(x)\equiv g(k(x)), \;\;\;\;\;\;\;\;\; \ls(x) \equiv \lambda(k(x))
\ee
for the gravitational couplings regarded functions of the logarithmic scale factor
(whence the subscript `s'). Then eq.(\ref{2.141}) for the relationship 
$H=H(a)$ translates to 
\be\label{5.10}
\frac{d}{dx} \UU(x) =-\unm \; (3+3w) \Big [1-\frac{\xi^2}{3}\lambda_s(x) \Big ]
\ee
Furthermore, upon differentiating (\ref{5.9}) and using $k(x) = \xi H(x)$, along with
(\ref{5.10}) and (\ref{5.1}) we obtain the following system for  $g_s$ and $\lambda_s$:
\begin{subequations}\label{5.11}
\ba\label{5.11a}
&&\frac{d }{dx} \gs (x)= -\frac{(3+3w)}{2} \; \Big [ 1-\frac{\xi^2}{3} \lambda_s(x) \Big ]
\; \beta_g \Big (g_s(x), \lambda_s(x) \Big)\\[2mm]
&&\frac{d }{dx} \ls(x)=-\frac{(3+3w)}{2} \; \Big [ 1-\frac{\xi^2}{3} \lambda_s(x) \Big ]
\; \beta_\lambda \Big (g_s (x), \ls(x) \Big)\label{5.11b}
\ea
\end{subequations}
The system (\ref{5.11a}) and (\ref{5.11b}) is closed: it can be integrated
in $x$ without involving $\UU$ or any other of the cosmological quantities.
Solving this system directly is numerically more stable than integrating with
respect to $k$ and substituting
\be\label{5.12}
k(x) = \xi \; H(x) =\xi \; \hT \; e^{\UU (x)}
\ee
into the result, in particular as $k(x)$ becomes functionally dependent on $\UU$
then.

Sometimes it is helpful to analyze the  system of RG and cosmological equations  
in still another way, namely by treating the dimensionless cosmological constant $\lambda$
as the independent parameter. Considered functions of $\lambda$, the quantities $g$, ${\cal U}$,
and $x$ are easily seen to obey the evolution equations
\ba
&&\frac{d}{d\lambda} g(\lambda) = 
\frac{\beta_g(g(\lambda), \lambda)}{\beta_\lambda(g(\lambda),\lambda)}\nonumber\\[2mm]
&&\frac{d}{d\lambda} {\cal U}(\lambda) = \frac{1}{\beta_\lambda (g(\lambda), \lambda)} \label{5.13}\\[2mm]
&&\frac{d}{d\lambda} x(\lambda) = -\frac{1}{2} \Big 
[ \Big ( 1-\frac{\xi^2}{3}\lambda\Big ) \beta_\lambda (g(\lambda), \lambda)\Big ]^{-1}\nonumber
\ea
Below we shall apply these equations to the IR branch of the cosmology. Here the relationship between
$g$, ${\cal U}$, $x$ and $\lambda$ is single valued, which of course is necessary for the method to work. 
\subsection{A trajectory with realistic parameter values}
Since through every point in the $g$-$\lambda$--plane there passes exactly one RG trajectory, 
we can specify a trajectory by specifying one of its points. Dealing with
type IIIa trajectories,   we shall use the turning point for this purpose.
The line of all  turning points $(\gat, \lat)$  is given by 
$\beta_\lambda(\gat, \lat)=0$, which yields the condition (\ref{1.15}) in the linear regime of
the GFP. Therefore a trajectory is uniquely specified by the 
$g$-coordinate of its turning point, $\gat$ : 
\be\label{5.20}
(\gat, \lat)=  \Big (\gat, \frac{\varphi_2}{2\pi}\gat \Big )
\ee
Having fixed $\gat$, we then integrate both upward and downward from the turning point, obtaining
the  UV- and IR-branch of the trajectory, respectively.

In the numerical calculations we shall use the value
\be\label{extre}
\gat = 10^{-60}
\ee
which is motivated by the order of magnitude estimates in the introduction where we
matched the linearized IIIa trajectories against the experimental data. Furthermore, in order
to fix the zero-point of the RG time $\tau \equiv \ln (k/\kat)$ we use the estimate
(\ref{1.22}) for the turning point scale: $\kat= 10^{-30} \mp$. As a result, 
\be\label{5.21}
\tau(k) =\ln (k/\mp) + 30 \; \ln(10)
\ee
In particular, recalling $H_0\approx 10^{-60} \mp$, 
\ba\nonumber
&&\tau(k=\mp) = +30\; \ln(10) \approx +69\\[2mm]
&& \tau(k=\kat)=0\label{5.22}\\[2mm]
&&\tau(k=H_0) = -30 \; \ln(10) \approx -69\nonumber
\ea

\begin{figure}[t]
\begin{center}
\includegraphics[width=16cm, height=11cm]{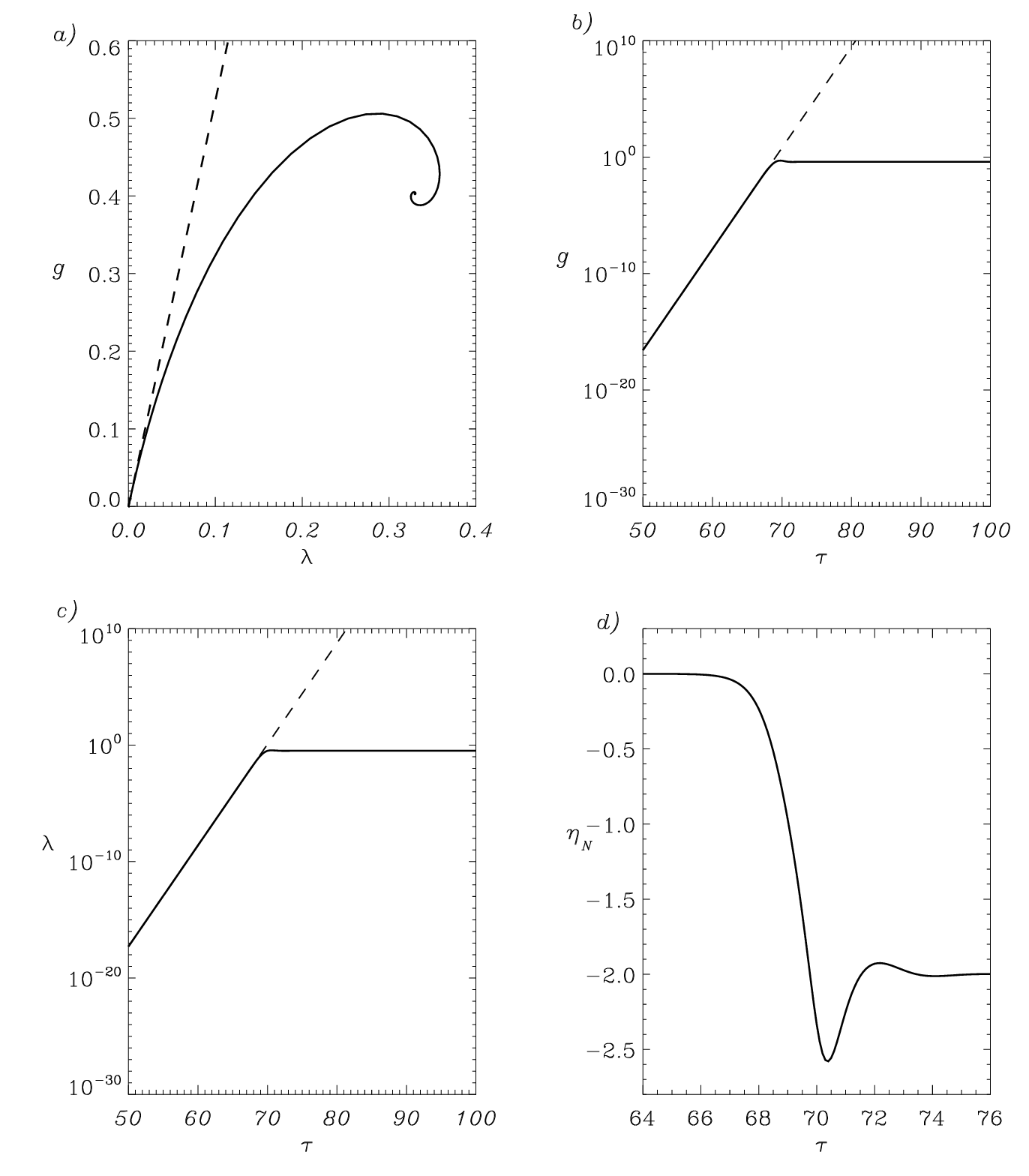}
\end{center}
\caption{ The realistic RG trajectory discussed in the text: Fig.~a) 
shows its shape on the $g$-$\lambda$--plane and compares it to the approximation
(\ref{1.20}) obtained by linearizing about the GFP (dashed line).
Figs. b) and c) show the scale dependence of $g$ and $\lambda$; the dashed
lines are the approximations from the GFP linearization. Fig.~d) displays the scale
dependence of the anomalous dimension.  } 
\label{fig3}
\end{figure}

Integrating the equations for $g(\tau)$, $\lambda(\tau)$ numerically
towards positive values of $\tau$ with the initial condition (\ref{5.20})
imposed at $\tau=0$ we obtain the UV branch of the trajectory approximately ``realized
in Nature''. It is displayed in Fig.~\ref{fig3} and compared to the approximation 
(\ref{1.20}) which we had obtained by linearizing about the GFP.

\begin{figure}[t]
\begin{center}
\includegraphics[width=15cm, height=11cm]{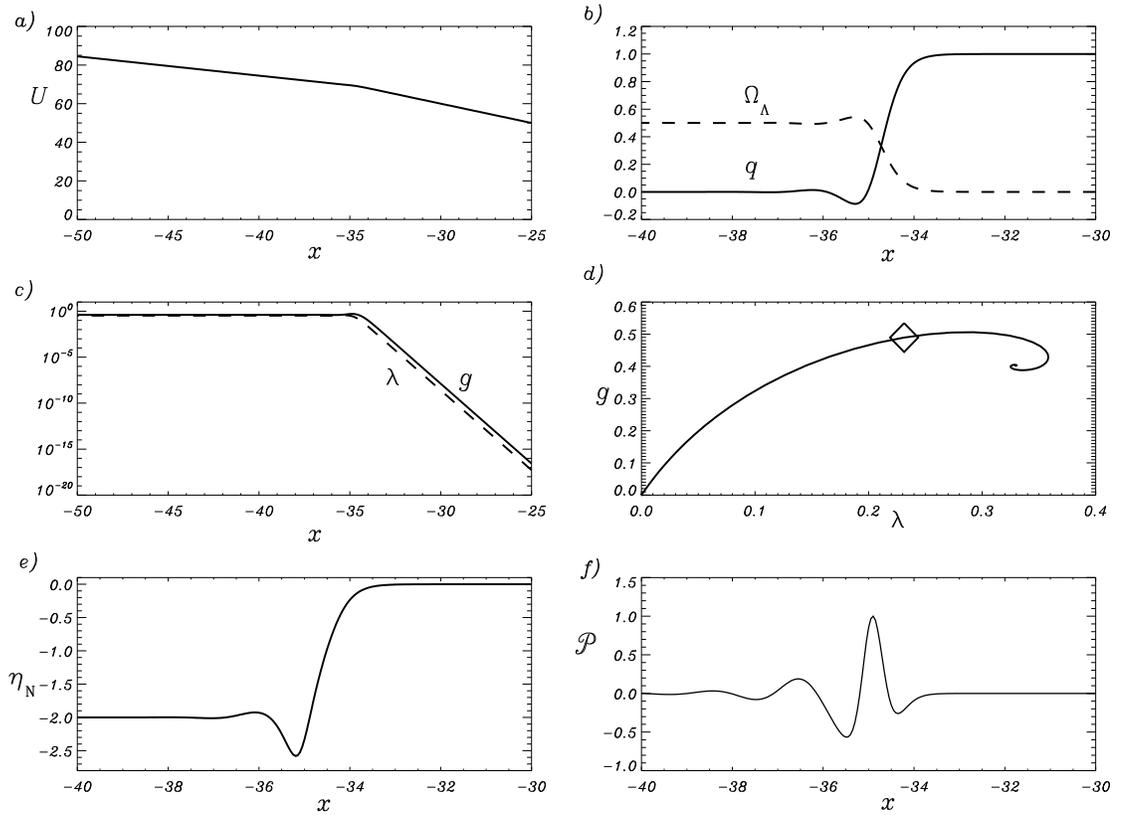}
\end{center}
\caption{  
The UV cosmology for $\OLS=0.5$. The plots a), b), c) display the logarithmic Hubble
parameter $\UU$, as well as $q$, $\OL$, $g$ and $\lambda$ as a function of the logarithmic
scale factor $x$. A crossover is observed near $x\approx -34.5$. 
The diamond in plot d) indicates the point on the RG trajectory corresponding to this $x$-value. 
The plots e) and f) show the $x$-dependence of the anomalous dimension and entropy
production rate, respectively}
\label{fig5}
\end{figure}

\begin{figure}[t]
\begin{center}
\includegraphics[width=15cm, height=11cm]{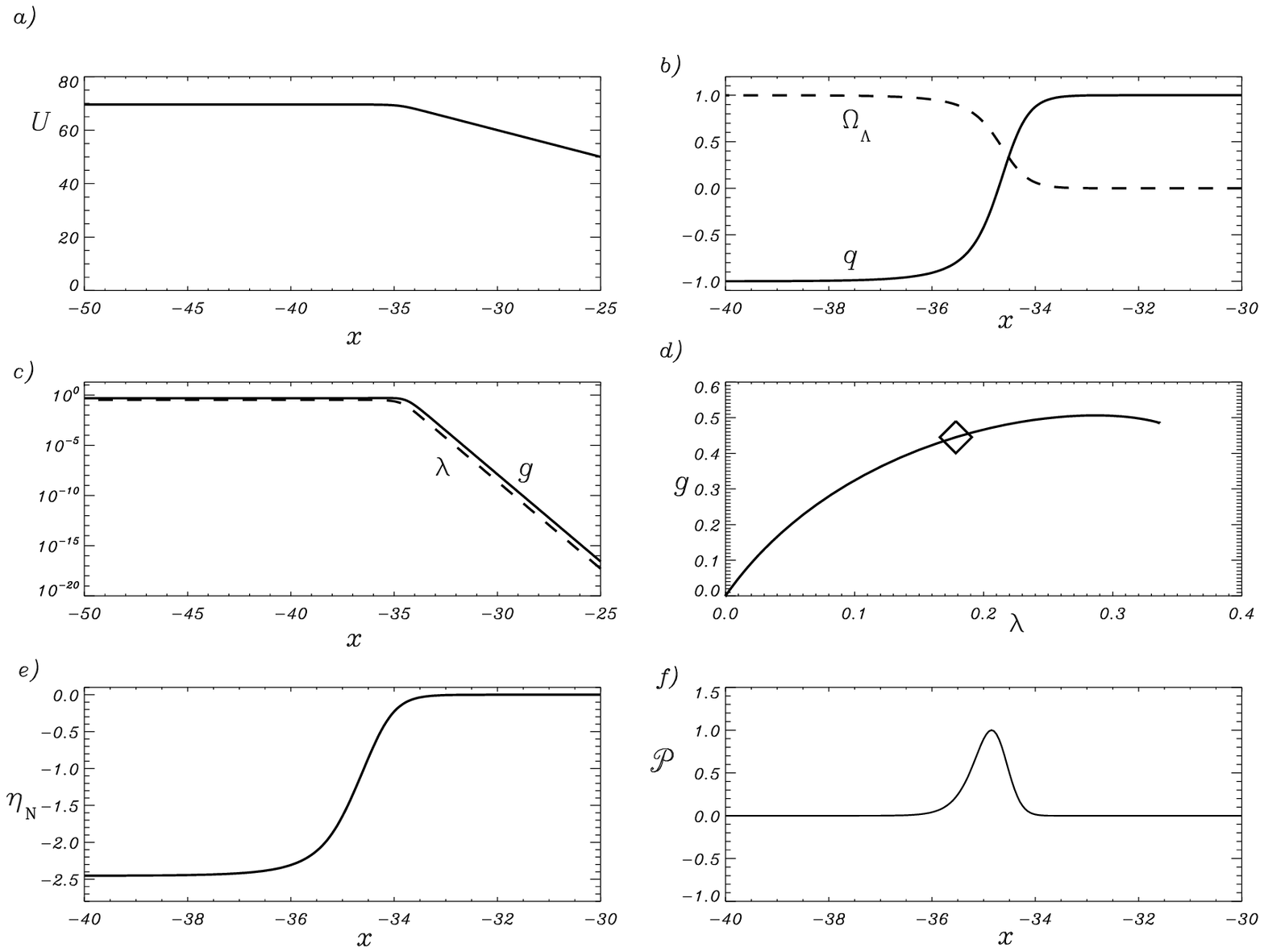}
\end{center}
\caption{  The same set of plots as in Fig.~\ref{fig5}, but for an UV cosmology with 
$\OLS= 0.98$. Note the almost vanishing slope of $\UU$ in the NGFP regime.}
\label{fig7}
\end{figure}

The plots of $g$ and $\lambda$ vs. $\tau$ show that for $\tau \lesssim 69$, 
\ie $k\lesssim \mp$, the linearization provides a reliable approximation while 
$g$ and $\lambda$ assume their constant fixed point values for $\tau \gtrsim 69$. The very
sharp bend in the $g(\tau)$  and $\lambda(\tau)$ curves indicates that the crossover from the
GFP to the NGFP is very rapid for the extreme initial condition (\ref{extre}). Because
of the extreme smallness of the beta functions, the trajectory spends a very long RG
time near the GFP and the NGFP, respectively. The transition from the linear scaling regime of 
the GFP to that of the NGFP happens in a short $\tau$-interval of the order $\Delta \tau\approx 6$.
This is more clearly seen in Fig.~3d) which shows the anomalous dimension plotted vs. $\tau$ .
It crosses over from $\eta_\ast ({\rm GFP}) =0$ to $\eta_\ast({\rm NGFP}) = -2$ between $\tau=66$
and $\tau = 72$, approximately.

Likewise one can obtain the IR branch of the trajectory by numerically integrating the equations
for $g(\tau)$, $\lambda(\tau)$ from $\tau=0$ towards negative values of $\tau$. We shall not 
display the plots here since the GFP linearization (\ref{1.20}) is a very precise approximation for most
values of $\tau < 0$. The equations can be integrated downward only to a finite termination scale 
$\tau_{\rm term} \equiv \ln (k_{\rm term}/k_{\rm T}) > - \infty$, i.e. $k_{\rm term} > 0$.
At this point $\eta_{\rm N}$ diverges to $-\infty$ and the $\beta$-functions  are undefined. For the
realistic initial conditions (\ref{5.20}) and (\ref{extre}) we find 
\be\label{5.23} 
\tau_{\rm term} \approx -69 \;\;\;  \Longrightarrow \; \; k_{\rm term} \approx H_0
\ee
where (\ref{5.22}) was used. Quite remarkably, and in accordance with the discussion in \cite{h3}, 
{\it the termination scale of the realistic trajectory is about the present Hubble parameter}.
At this scale  $\lambda$ has reached the value $1/2$, while $g$ is of the order $10^{-120}$ there.

As a consequence of this tiny $g$-value, the breakdown of the Einstein-Hilbert truncation near
$\tau_{\rm term}$ happens in a very abrupt way. In this regime the anomalous dimension (\ref{5.3}) is
well approximated by $\eta_{\rm N}\approx -(6/\pi) g (1-2\lambda)^{-1} \approx -10^{-120} (1-2\lambda)^{-1}$,
and this function jumps almost step-function like from $\eta_{\rm N}\approx 0$ at $\lambda <1/2$ to 
$\eta_{\rm N}=-\infty$ at $\lambda=1/2$. Thus, immediately before the termination, the trajectory
is still essentially classical: $\Lambda=const$, $G=const$. In fact, the numerics confirm the discussion in \cite{h3}:
In the IR branch the quantum effects die off already one or two orders of magnitude below 
$k_{\rm T}$ where then  a very long classical regime starts (between $\tau \approx -3$ and $\tau \approx -69$, say).
In this regime the GFP-linearization (\ref{1.20}) provides an excellent approximation. 
In order to determine the fate of the trajectory below $\tau_{\rm term}$ 
a better truncation would be needed.

\subsection{The UV branch of the cosmology}
In this subsection we describe the results obtained by numerically integrating the
eqs. (\ref{5.10}) and (\ref{5.11}) from $x=0$ towards large negative values. This amounts to 
going back in time, towards the UV, starting at the turning point. For the equation of state
parameter we choose $w=1/3$ which corresponds to radiation dominance.

After fixing the initial conditions (\ref{5.20}), (\ref{extre}) only one parameter remains to be
fixed, namely $\xi$, or equivalently the $\Omega_\Lambda$-value in the NGFP regime:
\be\label{5.30}
\xi^2 = (3/\lst) \; \OLS
\ee
Since $k$ is supposed to be of the order of $H$ we require $\xi=O(1)$. As $\lst=O(1)$
this is indeed the case if $\OLS=O(1)$. In principle $\OLS$ can vary over the full
interval $(0,1)$. For $\OLS$ ``anomalously'' close to zero, the condition $\OLS=O(1)$ is
violated however and we exclude such choices. On the other hand, $\OLS$-values very close to 1 are
perfectly allowed. We shall study the UV cosmology in dependence on $\OLS$ and are 
particularly interested in the limit $\OLS \nearrow 1$, \ie $\Omega_{\rm M} \searrow 0$.
(A possible dominance of the vacuum over the matter energy density would be nicely consistent
with the physical picture that part or all of the matter is generated by ``cosmological particle
production''  during the NGFP regime. )

It will be helpful to compare the exact numerical results to the predictions of the
classical FRW cosmology with $\Lambda=0$. This yields, for instance, $\UU(x) = \tau(x)=-2x$.
Since, by (\ref{5.22}), $k$ and hence $H$ are of the order $\mp$ for $\tau \approx 69$, 
the classical prediction for the logarithmic scale factor at which $k\approx H=O(\mp)$ is
\be\label{5.31}
x_{\rm FRW}^{\rm Pl} \approx -34.5
\ee 

Fig.~4 displays the result of the numerical solution for $\OLS=1/2$. The plots a), b), c) show
$\UU$, $\OL$, $\lambda$ and $g$ as a function of the logarithmic scale factor $x$. We observe
a crossover quite precisely at $x_{\rm FRW}^{\rm Pl}\approx -34.5$ . 
For smaller scale factors (earlier times)
the exact numerical solution is well approximated by the analytic fixed point solution (\ref{2.13})
with $w=1/3$   and $\OLS = 1/2$. It has $\alpha =1$, $\UU (x) = -x$, $\OL=const=0.5$, 
$q=0$, $(g,\lambda)= (\gst, \lst)=const$, in accord with the plots. For scale factors larger than 
$x_{\rm FRW}^{\rm Pl}$ the behavior is essentially that of a classical 
FRW cosmology without a substantial
cosmological constant: $\alpha = 1/2$, $\UU(x) = -2x$, $\OL=const\approx 0$, $q=1$. This 
explains why the crossover occurs almost exactly at the value (\ref{5.31}) predicted by the classical
$\Lambda=0$ theory. In Fig.~4d) we redisplay the RG trajectory on the $g$-$\lambda$--plane
and indicate by a diamond the point on the trajectory corresponding to $x=-34.5$. In the plots
4b) and 4e) we see that the ``width'' of the crossover is about 2 units of $x$ (``$e$-folds'').
In particular, the anomalous dimension changes from the canonical value 
$\eta_{\rm N}=0$ to the NGFP value 
$\eta_{\rm N}=-2$
between $x\approx -34$  and 
$x\approx -36$. From Fig.~4f) we learn that only in this interval the entropy production rate
$\clp$ is significantly different from zero. Note that $\clp$ is negative during short time
intervals; its time integral is positive though. 

Fig.~5  shows the analogous plots for $\OLS=0.98$. 
The crossover is found at the same scale $x_{\rm FRW}^{\rm Pl}\approx -34.5$. The cosmology for 
$x\gtrsim x_{\rm FRW}^{\rm Pl}$, for any value of $\OLS$ in fact, is essentially the classical 
$\Lambda=0$ cosmology again. The numerical results for $x< x^{\rm Pl}_{\rm FRW}$ approach
the analytic fixed point solution with an exponent $\alpha = (2-2\OLS)^{-1} >1$ for
$\OLS >0.5$ corresponding to a ``power law inflation'' $a(t) \propto t^\alpha$. Consistent
with that we see that when $\OLS \nearrow 1$ the slope of $\UU (x) = -2 (1-\OLS)x$ decreases and
finally vanishes at $\OLS=1$. This limiting case corresponds  to a constant Hubble parameter, 
i.e. to de Sitter space. For values of $\OLS$ smaller than, but close to $1$ this 
de Sitter limit is approximated by an expansion $a\propto t^\alpha$ with a very large
exponent $\alpha$. We can see this trend when we compare the plots a) of Figs.~4 and 5, 
respectively. In Fig.~5, the logarithmic Hubble parameter has almost no visible $x$-dependence in the
NGFP regime. We shall come back to this power law inflation in more detail later on.

Another feature which distinguishes the $\OLS> 1/2$ cosmologies from the case $\OLS=1/2$
is that entropy is produced in the NGFP regime, see subsection 4.1. The entropy production 
rate $\clp$ is plotted in Fig.~4f) and 5f), respectively. The contribution from 
the NGFP regime is not visible in Fig.~5f) , however, since there $\clp$  is much smaller
than at the peak in the crossover region.

Summarizing the numerical results we can say that for any value of $\OLS$ the UV cosmologies
consist of two scaling regimes with a relatively sharp crossover region near
$k,H\approx \mp$ which separates them. At higher $k$-scales the fixed point approximation 
(\ref{4.13}) is valid, at lower scales one has a classical FRW cosmology in which $\Lambda$
can be neglected. The $k^4$-cosmology discussed analytically in subsection 4.2 would be
valid near the crossover, but it seems not to be realized for a significant number
of $e$-folds. 

We have not yet related the (logarithmic) cosmological time $y$ to the scale factor $x$. In principle
the function $y=y(x)$ could be obtained by integrating $d t (a) /da = [a H(a) ]^{-1}$.
We shall not need the exact relationship here. Since FRW cosmology is valid for $t \gtrsim \tp$ 
the classical relation $t\propto a^2$ or $y=2x$ is an excellent approximation for all $t\gtrsim \tp$.

\subsection{The IR branch of the cosmology}
By integrating the improved cosmological  equations from $x=0$ towards positive values of $x$
we obtain a $1$-parameter family of cosmologies which are valid after the turning point
of the RG trajectory has been passed. The free parameter is $\xi=k/H$. In the UV 
cosmology we used (\ref{5.30}) in order to express $\xi$ in terms of the more physical parameter
$\OLS$. If $\xi$ was strictly constant
all the way from the Planck regime to asymptotically late times then we could keep 
using (\ref{5.30}) in the IR, of course. However, the cutoff identification $k=\xi H$ is only an 
approximation. Hence  it would be unrealistic to assume that, in case $k$ is always approximately proportional to $H$, the constant 
of proportionality is strictly time independent. For this reason we allow $\xi$ in the late Universe
to be different from its value in  the very early Universe. So the parameter $\xi$ labeling the 
different IR cosmologies does not necessarily satisfy (\ref{5.30}). 

In subsection 2.3 we saw that the ``$\Omega$-line'' along which $\rho=0$ is a 
straight line on the $g$-$\lambda$--theory space, parallel to the $g$-axis at 
$\lambda_\Omega=3/\xi^2$. Depending on $\xi$, the $\Omega$-line can be within the domain
of validity of the Einstein-Hilbert truncation ($\lambda_\Omega \lesssim 1/2$) or outside 
($\lambda \geq 1/2$). Only in the first case the RG improved field equations possess solutions which 
realize what was called the ``$\Omega$-mechanism'' in [II].

Consider a solution with $da/dt>0$ at late times describing a Universe which keeps expanding
for ever, i.e.  there is no recontraction.  Hence its matter contents gets continously diluted
(at least in absence of particle creation) and at asymptotically late times one has
$\rho (t\rightarrow \infty)=0$. In the first case above this entails that, for $t\rightarrow \infty$,
the Universe is described by a pair $(g,\lambda)$  on the $\Omega$-line. 
Remarkably, if $\lambda_\Omega < 1/2$, any RG trajectory of type IIIa hits the $\Omega$-line at a 
non-zero value of $k$, see Fig.~2.  As a result, the asymptotically late Universe is characterized by a 
{\it constant} and {\it non-zero} scale:
\be\label{5.32}
k_{\rm asym}= k(t\rightarrow \infty)>0
\ee
During its entire history the Universe does not probe the complete RG trajectory, but only
the portion with $k> k_{\rm asym}$. If $\lambda(k_{\rm asym})=\lambda_\Omega$ is still sufficiently
far below $1/2$, the Einstein-Hilbert truncation can describe the latest stages of 
the cosmological evolution even. In fact, since its breakdown happens very abruptly near $\lambda=1/2$
and before that $\eta_{\rm N}$ is almost zero, we see that, if the $\Omega$-mechanism takes
place, the late cosmology is essentially classical; no significant renormalization effects occur.

With the cutoff identification $k = \xi H$ adopted in this paper the interpretation of the
asymptotic regime with $k=k_{\rm asym}=const$ is clear\footnote{In [II] the situation was
slightly more complicated since a dynamical cutoff identification had been used. } : it amounts
to an asymptotic de Sitter phase with a constant Hubble parameter $H_{\rm asym}= k_{\rm asym}/\xi$ .
Or, using  (\ref{2.5a}) for $\rho=0$, 
\be\label{5.33}
H_{\rm asym}= \sqrt{\Lambda (k_{\rm asym})/3}
\ee
Since $k_{\rm asym} \ll k_{\rm T}$ we may use (\ref{1.16}) to rewrite (\ref{5.33}) as 
\be\label{5.34}
H_{\rm asym}= \sqrt{ \Lambda_0 /3} = k_{\rm T} \sqrt{\lambda_{\rm T} /6}
\ee
Neglecting factors of order unity, this relation yields the asymptotic $\UU$ value
\be\label{5.35}
\UU_{\rm asym}\approx \frac{1}{2} \ln (\lambda_{\rm T})
\ee
For the realistic trajectory with the initial conditions  (\ref{5.20}), (\ref{extre})
we have $\UU_{\rm asym} \approx -30 \ln (10) \approx -69$. 

For a numerical investigation of the $\Omega$-mechanism it is most convenient to integrate the 
evolution equations with respect to $\lambda$ as the independent parameter, see eqs.(\ref{5.13}) .
In Fig.~6 we show the results for the example with $\xi=2.86$ which has $\lambda_\Omega=0.367$.
(If we insist on a strictly constant $\xi$ this would correspond to $\OLS = 0.90$ .)
\begin{figure}[t]
\begin{center}
\includegraphics[width=15cm, height=11cm]{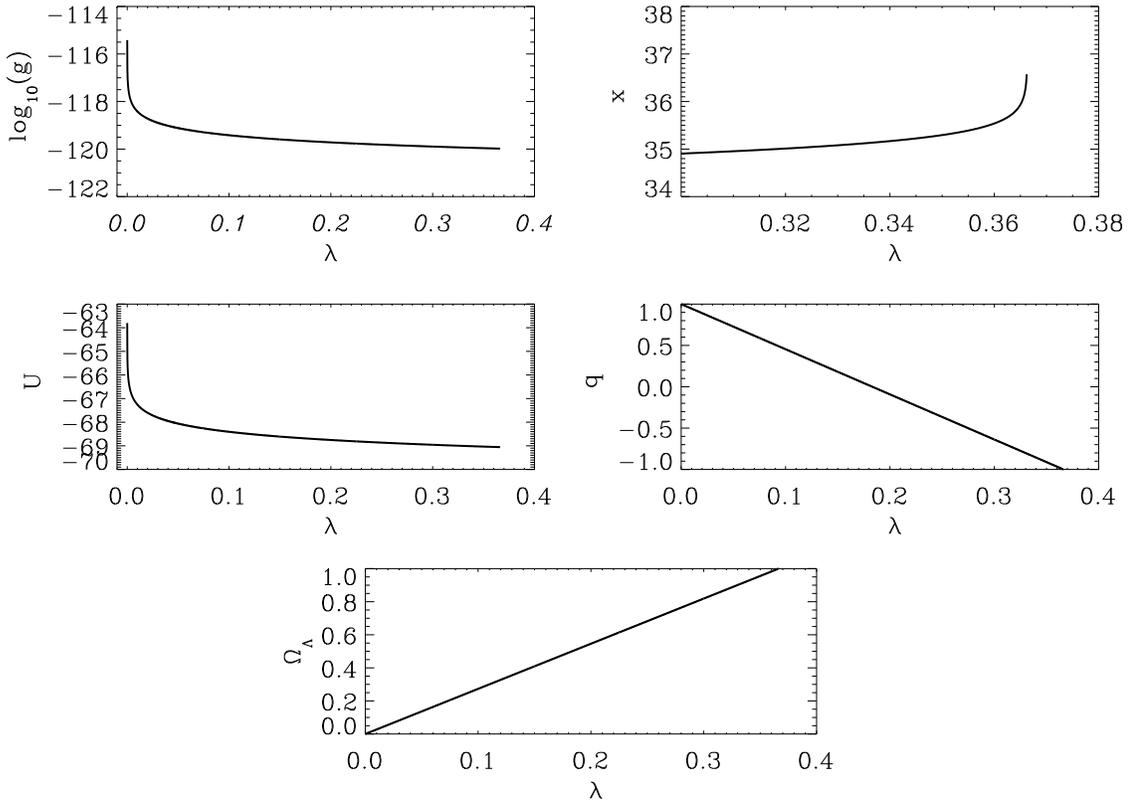}
\end{center}
\caption{ Example of an IR cosmology with $\Omega$-mechanism. The plots show
$g$, $\UU$, $q$ and $\Omega_\Lambda$ as a function of the dimensionless cosmological
constant $\lambda$. Note in particular that when $\lambda$ approaches
$\lambda_\Omega \approx 0.367$ the scale factor grows rapidly, and $q$ and 
$\Omega_\Lambda$ approach their de Sitter values $-1$ and $+1$, respectively.}
\label{fig8}
\end{figure}

The Fig.~6b)  shows that $\UU$ indeed approaches the constant value $-69$ asymptotically.
The plot in Fig.~6c) displays the logarithmic  scale factor as a function of $\lambda$.
We see that $x$ diverges for $\lambda \nearrow \lambda_\Omega$, which is precisely the
signature of the $\Omega$-mechanism. The Fig.~6d) and 6e) confirm that the cosmology directly 
at $\lambda=\lambda_\Omega$ is an almost ``empty'' de Sitter Universe with $q=-1$ 
and $\Omega_\Lambda=1$. The energy density of ordinary matter has dropped to zero there. 
One can check that this entire cosmology is essentially classical. The quantum corrections to 
the beta-functions are negligible on the entire interval 
$\lambda_{\rm T} \leq \lambda \leq  \lambda_\Omega$, which implies in particular that there
is no entropy production.

This calculation has been performed for $w=1/3$. To be more realistic one should compute
the last few $e$-folds with $w=0$ corresponding to matter dominance. This will not
change the overall conclusion, however, that there are no significant quantum effects in the
IR cosmologies with $\Omega$-mechanism. As long as $\lambda_\Omega < 1/2$, every value
of $\xi$ yields essentially the same cosmology.

As a result, we may use the classical FRW formulas to estimate the time when the 
Universe starts accelerating due to the transition from matter or radiation dominance
to $\Lambda$-dominance. For a first orientation we can use (\ref{5.7}) to estimate $x$
and $y$ at the onset of the acceleration:
\ba\label{5.36}
&&x_{\rm acc} \approx -\UU_{\rm asym} /2 \approx 34.5\\[2mm]
&&y_{\rm acc} \approx -\UU_{\rm asym} \approx 69
\ea
Interestingly enough, these numbers correspond roughly to the scale factor and age of the Universe
we live in. In particular its age is 
$t_0\approx t_{\rm acc} \approx 10^{30} \; t_{\rm T} \approx 10^{60} \; t_{\rm Pl}$.

It is important to understand what determines the time $t_{\rm acc}$ at which the Universe
switches from deceleration to acceleration as the vacuum energy starts dominating the matter
energy density. According to the observations, this has happened only ``very recently''
in our cosmological history, so the natural and frequently posed question is ``why just now?''
\cite{now,coscon}.

In the present setting the answer to this question is clear: $t_{\rm acc}$ is determined
by the asymptotic value of the cosmological constant which, in turn, is dictated
by the RG trajectory. Hence, $t_{\rm acc}$ is what it is because Nature's RG trajectory
is what it is. 

This might appear to be a rather tautological statement at first sight, 
in particular since we actually used the observed cosmological constant to fix the parameters of
the trajectory. However, QEG  is believed to be a predictive theory \cite{oliver3,livrev}
in the sense that at the exact level only {\it finitely many} parameters need to  be taken
from the experiment in order to completely determine the trajectory, and that
then {\it infinitely many} predictions are possible in terms of those\footnote{ See \cite{livrev}
for a more precise discussion of this point.}. One of the input parameters is $\Lambda$
at some scale, and there are {\it a few} more such input parameters. One of the predictions
is $t_{\rm acc}$, but there are {\it infinitely many} more such predictions. 

It is only because
of the observational situation we are in that the above statement about $t_{\rm acc}$ appears
tautological. Since the only determination of $\Lambda$ which is available to date is 
on cosmological scales, we are 
forced to use this cosmological $\Lambda$ as an input parameter and therefore cannot predict
$\Lambda (k_{\rm asym})$ or $t_{\rm acc}$ in terms of anything independent. Instead, 
we are able to make  ``predictions'' about the early Universe in terms of the parameters fixed in
the late Universe. Conceptually the situation is the same as in 
(perturbative) QED for instance. The pertinent RG trajectories have 2 free parameters. It is
convenient but by no means compulsory to choose them as the mass $m$ and charge $e$
of the electron. If one does so, $e$ and $m$ are no predictions of course, but if instead we
parametrize the trajectory by two different couplings $g_1 \equiv g_1 (e,m)$ and 
$g_2\equiv g_2 (e,m)$ and measure $g_1$ and $g_2$ then $e$ and $m$ are predicted by the 
theory in terms of $g_1$ and $g_2$. Likewise we can imagine a (considerably) improved
experimental situation in which all the parameters of the gravitational RG trajectory can 
be determined from laboratory measurements. The time $t_{\rm acc}$ and similar cosmological 
quantities are true predictions then.

If $\lambda_\Omega > 1/2$ there is no $\Omega$-line which would prevent the RG
trajectory underlying the cosmological evolution to run into the singularity at 
$\lambda = 1/2$. In this case the coupled RG and cosmological equations cannot be integrated
beyond a certain point where the Einstein-Hilbert truncation breaks down. As we explained in 
the Introduction already, this breakdown, if it occurs, is expected to happen when
$k\approx H_0$ , i.e. ``just now''. For smaller $k$-scales one would have to use a 
more general truncation of  theory space whose implications for the cosmology
cannot be guessed offhand. A theoretically attractive (and phenomenologically viable)
possibility could be the IR fixed point model of refs.\cite{cosmo2,elo} .

\section{Inflation in the fixed point regime}
\renewcommand{\theequation}{6.\arabic{equation}}
\setcounter{equation}{0}
In this section we discuss in some detail the epoch of power law inflation which is realized
in the NGFP regime if $\OLS > 1/2$. Since, as we saw in the previous section, 
the transition from the fixed point to the classical FRW regime is rather sharp it will be
sufficient to approximate the RG improved UV cosmologies by the following caricature :
For $0<t< t_{\rm tr}$, where $t_{\rm tr}$ is a transition time, the scale factor behaves as
$a(t)\propto t^\alpha$, $\alpha > 1$.  Here $\alpha = (2-2\OLS)^{-1}$ since $w =1/3$ 
will be assumed. Thereafter, for $t>t_{\rm tr}$, we have a classical, entirely
matter-driven expansion $a(t)\propto t^{1/2}$ . 
\subsection{Transition time and apparent initial singularity}
The transition time $t_{\rm tr}$ is dictated by the RG trajectory. 
The latter leaves the asymptotic scaling 
regime near $k\approx \mp$. Hence $t_{\rm tr}$ is the time at which 
$k(t_{\rm tr})=\xi H(t_{\rm tr})\approx \mp$.
In the following we only consider values of $\OLS$ in the interval 
$(1/2, 1)$ because there is no inflation
otherwise. For such values of $\OLS$, and since $\lambda_\ast = O(1)$, eq.(\ref{4.110}) tells us that 
\be\label{6.1}
\xi=\sqrt{3\OLS/\lambda_\ast}
\ee
is of order unity so that we can determine $t_{\rm tr}$ 
from $H(t_{\rm tr})\approx \mp$. Using (\ref{4.11})
at the matching point we find 
\be\label{6.2}
t_{\rm tr}= \alpha \; \tp
\ee

This is an important relation and several comments are in order here. Let us recall that, 
as always in this paper, the Planck mass, time, and length are defined in terms of the value of Newton's
constant in the classical regime, cf. the discussion following eq.(\ref{1.21}) :
\be\label{6.3}
\tp = \lp = \mp^{-1} = \bar{G}^{1/2} = G_{\rm observed}^{1/2}
\ee
For the sake of the argument, let us now assume that $\OLS$ is very close 
to $1$ so that $\alpha$ is large:
$\alpha \gg 1$. Then (\ref{6.2}) implies that the transition takes place at a cosmological time 
which is much later 
than the Planck time. At the transition the {\it Hubble parameter} is of order $\mp$, but the 
{\it cosmological time } is in general not of the order of $\tp$. Stated differently, the ``Planck time''
is {\it not} the time at which $H$ and the related physical quantities assume Planckian values. 
Turning (\ref{6.2}) around we conclude that the Planck time as defined above is well within the NGFP regime:
$\tp = t_{\rm tr} / \alpha \ll t_{\rm tr}$. 

At $t=t_{\rm tr}$ the NGFP solution (\ref{4.13}) is to be matched continuously with a FRW cosmology
(with vanishing cosmological constant ). We may use the familiar classical formulas such as (\ref{3.31})
for the scale factor, but we must shift the time axis on the classical side such that $a$, 
$H$, and then as a result of (\ref{2.5a}) also $\rho$ are continuous at $t_{\rm tr}$. Therefore
$a(t)\propto (t-t_{\rm as})^{1/2}$ and
\be\label{6.30}
H(t) = \frac{1}{2} \; (t-t_{\rm as})^{-1} \;\;\; \text{for } \;\;\; t> t_{\rm tr}
\ee
Equating the Hubble parameter (\ref{6.30}) at $t=t_{\rm tr}$ to 
$H(t) = \alpha /t$, valid in the NGFP regime, we find that the shift $t_{\rm as}$ must be chosen
as 
\be\label{6.31}
t_{\rm as} = \Big (\alpha -\frac{1}{2} \Big ) \; \tp = 
\Big (1 - \frac{1}{2\alpha}\Big ) \; t_{\rm tr} \; < \; t_{\rm tr} 
\ee
Here the subscript 'as' stands for ``apparent singularity''. This is to indicate that if one continues
the classical cosmology to times $t<t_{\rm tr}$, it has an initial singularity (``big bang'') at 
$t=t_{\rm as}$. Since, however, the FRW solution is not valid there nothing special happens at 
$t_{\rm as}$; the true initial singularity is located at $t=0$ in the NGFP regime. (See Fig.~7.)

We emphasize that for any choice of $\OLS$, and hence $\alpha$, one always has 
\be\label{6.32}
H(t_{\rm tr})=\mp
\ee
At the moment when the classical cosmology starts 
becoming valid, whatever was its ``prehistory'', it starts
with $H\approx \mp$ and $\rho \approx \mp^4$. 

\subsection{Crossing the Hubble radius}
In the NGFP regime $0<t< t_{\rm tr}$ the Hubble radius $\ell_{H} (t) \equiv 1/H(t)$, i.e.
\be\label{6.4}
\ell_{H} (t) = \frac{1}{\alpha} \; t , 
\ee
increases linearly with time but, for $\alpha \gg 1$, with a very small slope. At the transition, the slope 
jumps from $1/\alpha$ to the value $2$ since $H=1/(2t)$ 
and $\lp=2t$ in the  FRW regime. This behavior is sketched in Fig.~7. The length scale $\LH$ 
measures the radius of curvature of spacetime. It has no interpretation as the distance
to a horizon: Robertson-Walker spacetimes with $a(t\rightarrow 0)\propto t^\alpha$, $\alpha > 1$,
have no particle horizon. At the transition time
$\LH(\tr)=\lp$.

Let us consider some structure of comoving length $\Delta x$, 
a single wavelength of a density perturbation,
for instance. The corresponding physical, i.e. proper length is $L(t) = a(t) \Delta x$ then. 
In the NGFP regime it has the time dependence 
\be\label{6.5}
L(t) = \Big (\frac{t}{\tr} \Big )^{\alpha} \; L(\tr)
\ee 
The ratio of $L(t)$ and the Hubble radius evolves according to 
\be\label{6.6}
\frac{L(t)}{\LH (t)} = \Big ( \frac{t}{\tr} \Big )^{\alpha-1} \; \frac{L(\tr)}{\LH (\tr)}
\ee
For $\alpha > 1$, i.e. $\OLS > 1/2$, the proper length of any object grows faster than the Hubble
radius. So objects which are of ``sub-Hubble'' size at early times can cross the Hubble radius and become
``super-Hubble'' at later times, see Fig.~7. 

Let us focus on a structure which, at $t=\tr$, is
$e^N$ times larger than the Hubble radius. Before the transition we have
\be\label{6.7}
L(t)/\LH (t) = e^N \; (t/\tr)^{\alpha -1}
\ee
Assuming $e^N > 1$, there exists a time $t_N < \tr$ at which $L(t_{N}) =\LH (t_{ N})$
so that the structure considered ``crosses'' the Hubble radius at the time $t_N$. Using (\ref{4.180}) it is given
by 
\be\label{6.8}
t_{N}=\tr \;{\rm exp}  {\Big ( -\frac{N}{\alpha -1} \Big )} = \tr \; {\rm exp} 
\Big [ -\frac{(1-\OLS) N}{(\OLS-1/2)} \Big ]
\ee
What is remarkable about this result is that, even with rather moderate values of $\alpha$, one can 
easily ``inflate'' structures to a size which is by many $e$-folds larger than the Hubble radius 
{\it during a very short time interval at the end of the NGFP epoch}. 

Let us illustrate this phenomenon by means of an example, namely the choice $\OLS = 0.98$ used
in Fig.~5.
Corresponding to $98\%$ vacuum and $2\%$ matter energy density in the NGFP regime, this value
is  still ``generic'' in the sense that $\OLS$ is not fine tuned to equal unity 
with a precision of many decimal places. It leads to the exponent $\alpha = 25$, the transition
time $\tr = 25 \; \tp$, and $t_{\rm as}=24.5 \; \tp$. 

The largest structures in the present Universe, evolved backward in time by the classical equations
to the point where $H=\mp$, have a size of about $e^{60}\; \lp$ there. We can use 
(\ref{6.8}) with $N=60$ to find the time $t_{\rm 60}$ at which those structures crossed
the Hubble radius. With $\alpha = 25$ the result is $t_{\rm 60}=2.05\; \tp = \tr /12.2$. Remarkably, $t_{\rm 60}$ is smaller than 
$\tr$ by one order of magnitude only. As a consequence, the physical conditions prevailing at the
time of the crossing are not overly  ``exotic'' yet. 
The Hubble parameter, for instance, is only one order of magnitude larger than at the transition:
$H(t_{\rm 60})\approx 12 \mp$.  The same is true for the temperature; eq.(\ref{4.230}) implies 
$T(t_{\rm 60})\approx 12 T(\tr)$ where $T(\tr)$ is of the order of $\mp$. Note also that $t_{\rm 60}$ is larger than $\tp$.
\subsection{Primordial density fluctuations}
QEG offers a natural mechanism for generating primordial fluctuations during the NGFP epoch which have a
scale free spectrum with a spectral index close to $n=1$. This mechanism is at the very heart of the 
``asymptotic safety'' underlying the nonperturbative renormalizability of QEG. It might open an observational window
which allows us a view of the gravitational physics in a regime where we expect qualitatively 
important quantum effects. Hence this issue could be of interest for the program of asymptotic safety per se
and not only for cosmology. 

The cosmology of the very early Universe reflects properties  of the RG trajectory close to the fixed point. 
In this regime the anomalous dimension of the graviton is very close to $\eta_{\rm N}^\ast = -2$, its value directly
at the  NGFP (in $d=4$). 

Using the effective field theory properties of $\Gamma_k$ it was shown in \cite{oliver1} that the graviton propagator
implied by the standard effective action $\Gamma_{k\rightarrow 0}$, on a flat background, has a large 
momentum behavior $\widetilde {\cal G}(p) \propto 1/p^4 $ which amounts to ${\cal G}(x;y) \propto {\rm ln} (x-y) ^2$
in position space. This form of the propagator is valid for $p^2\gg \mp^2$ and $(x-y)^2\ll \lp^2$, respectively. 
It is a direct consequence of $\eta_{\rm N}^\ast = -2$, In fact, the logarithmic position dependence can be
understood as a limiting case of the standard critical 2-point function ${\cal G}(x;y) \propto 1/|x-y|^{d-2+\eta}$
for $d=4$ and $\eta \rightarrow -2$. 

Following [I], let us now consider curvature fluctuations $\delta {\bf R}$ caused by metric fluctuations $h_{\mu\nu}(x)$.
In a symbolic notation we have $\delta {\bf R} \propto \partial \partial h$ where ${\bf R}$ stands for any component of the Riemann
tensor. As $\langle h_{\mu\nu}(x) h_{\rho\sigma}(y) \rangle \propto \ln(x-y)^2$ the 2-point function of $\delta {\bf R}$
is found to be $\langle \delta {\bf R}(x) \delta {\bf R}(y)\rangle \propto 1/(x-y)^4$. 
(In the classical regime we would have a decay $\propto 1/(x-y)^6$ instead.) Up to now the background was assumed flat.
Allowing for a curved background spacetime, the above formulae will give the leading short distance behavior:
\be\label{6.9}
{\cal G}(x;y) \propto \ln \; d(x,y)^2, \; \; \; \langle \delta {\bf R}(x) \delta {\bf R}(y)\rangle \propto \frac{1}{d(x,y)^4}
\ee
Here $d(x,y)$ is the geodesic distance of $x$ and $y$. Eqs. (\ref{6.9}) are valid provided $d(x,y)$
is smaller than the radius of  curvature of the background spacetime (and $\lp$, of course). 
In a Robertson-Walker geometry this condition amounts to $d(x,y)< \LH(t)$, i.e. (\ref{6.9}) is valid
on ``sub-Hubble'' scales. 

Next assume $x$ and $y$ are two points on the same time slice of a Robertson-Walker spacetime. 
Their distance is $d(x,y) = a(t) \; |{\bf x} -{\bf y}|$ where ${\bf x}$ and ${\bf y}$ are the comoving Cartesian
coordinates of $x$ and $y$. Ignoring the time dependence, (\ref{6.9}) yields
\be\label{6.10a}
\langle \delta {\bf R}({\bf x}, t) \delta {\bf R}({\bf y}, t) \rangle \propto \frac{1}{|{\bf x}-{\bf y}|^4}
\ee
The above general arguments imply that these relations should be valid if $a(t) |{\bf x}-{\bf y}| \ll \LH (t) \ll \lp$. 
(At larger distances the 2-point function can be determined by a detailed computation only which has
not been performed yet.)
In the improved cosmologies with inflation we found that for any value of $\OLS$ the inequality 
$\LH (t) < \lp$ is satisfied for all $t< \tr$. Hence (\ref{6.10a}) is applicable, on sub-Hubble distances, during the 
entire NGFP era.

Let us now come back to the problem of  primordial density perturbations which could act as seeds for structure formation in
the Universe. Here we adopt the same hypothesis as in the standard inflationary scenarios \cite{lily,padstr}, namely
that they stem from quantum fluctuations which have effectively become classical. In models of scalar-driven
inflation it is usually the fluctuations of the ``inflaton'' itself which serves this purpose. In our case
inflation happens automatically as a consequence of the RG running and no inflaton is needed. 
Instead, it is the fluctuations of the geometry itself, i.e. of the metric and its curvature, which are the natural 
candidates for the seeds of structure formation. 

A quantity we have observational access to is the classical correlator of density perturbations, 
\be\label{6.11}
\xi({\bf x}) = \langle \delta ({\bf x}+{\bf y}) \delta ({\bf y})\rangle ,
\ee
where $\delta ({\bf x}) \equiv \delta \rho ({\bf x})/\rho$. 
If its power spectrum at a fixed instant of time, 
\be\label{6.12}
|\delta_k|^2 \equiv V \; \int dx^3 \; \xi ( {\bf x} ) \; \exp{(-i{\bf k}\cdot {\bf x})} , 
\ee
behaves as $|\delta_k|^2\propto |{\bf k}|^n$ the spectrum is said to have the spectral index $n$. 
From the observation of the CMBR we know that the perturbations $\delta \rho$ which got 
imprinted in the microwave background at decoupling had
an almost scale free (Harrison-Zeldovich) spectrum with $n\approx 1$. 

Remarkably, this is exactly the spectrum one obtains if the seeds of the density perturbations are sub-Hubble
fluctuations in the NGFP era. The reasoning in [I] was as follows. Already at the level of the classical 
Einstein equations,  density fluctuations $\delta  \rho$ are proportional to fluctuations $\delta {G_0}^0$ of the
Einstein tensor ${G_\mu}^\nu$, i.e. a special combination of $\delta {\bf R}$-components. Therefore, if fluctuations
of the geometry are the source of the density fluctuations, the correlators of $\delta\rho$ should at least approximatively
be proportional to that of $\delta {\bf R}$ as given in  eq.(\ref{6.10a}): $\xi({\bf x})\propto 1/|{\bf x}|^4$. Taking the 
Fourier transform one finds  $|\delta_k|^2 \propto |{\bf k}|$, i.e. the spectral index $n=1$. This argument is similar
in spirit to the discussion in \cite{antonmazmo}. It suggests that near $\tr$, when the Universe 
has become classical, density perturbations have been created from quantum fluctuations with a nearly scale
free spectrum, $n\approx 1$. 
\begin{figure}[t]
\begin{center}
\includegraphics[width=11cm, height=8cm]{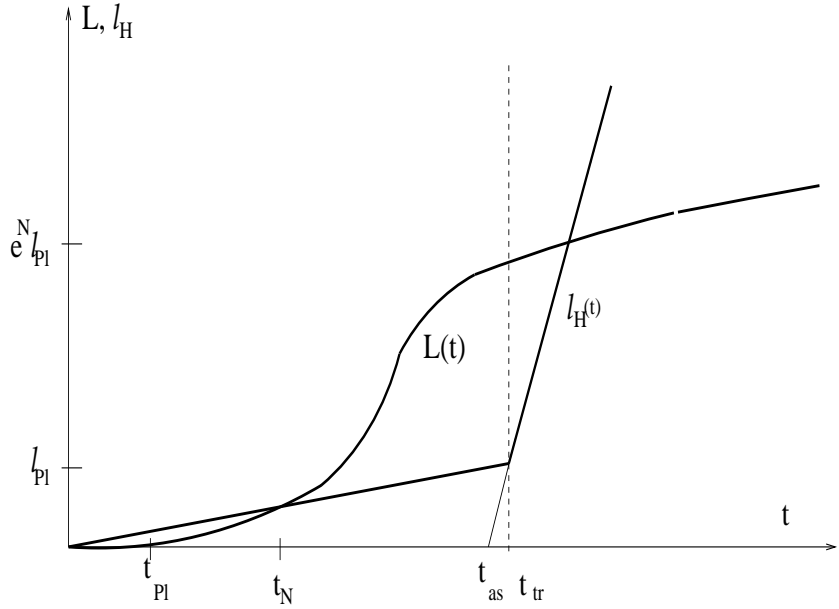}
\end{center}
\caption{Shown is the proper lenth $L$ and the Hubble radius as a function of time. 
The NGFP and FRW cosmologies are valid for $t<t_{\rm tr}$ and $t>t_{\rm tr}$, respectively.
The classical cosmology has an apparent initial singularity at $t_{as}$ outside its domain of 
validity. Structures of size $e^N \lp$ at $t_{\rm tr}$ cross the Hubble radius at $t_N$, a 
time which can be larger than the Planck time. }
\label{fig9}
\end{figure}

Since the evolution of the perturbations after $\tr$ is essentially classical we know that, to be of phenomenological relevance, the 
$n=1$ spectrum thus generated should apply to fluctuation modes with wave lengths as large as $\approx e^{60}\lp$, say, at $t=\tr$.
If the wavelength is larger than $\lp$ the scale free correlator (\ref{6.10a}) is not valid, most probably. 
However, in any of the cosmologies with $\OLS> 1/2$ there is an inflationary NGFP era.  As a consequence,  there exists always
a time $t_{\rm 60}$ before which the modes are completely within the Hubble radius and the above argument does apply.
(See Fig.~7.) If they  keep their spectrum during the expansion to super-Hubble scales we end up with a $n\approx 1$
spectrum at $t=\tr$. 

If $\alpha$ is large the crossing times $t_{ N}$  of all modes relevant to structure formation are close to $\tr$.
Hence they all become ``super-Hubble'' at about the same value of $H$ and $\LH$.

At this point there is a clear difference between the RG improved cosmology found in [I] by imposing the ``consistency
condition'' and the new ones found in the present paper. The cosmology of [I] has $a(t) \propto t$, 
i.e. no inflation. Therefore the above argument applies only to the modes which were sub-Hubble at $t_H = \tp$ and are
of millimeter size today.

\subsection{No reheating is necessary}
By combining (\ref{4.230}) with (\ref{4.21}) and (\ref{6.2}) we obtain the following expression for the 
temperature at the end of the NGFP regime:
\be\label{6.20}
T(\tr) = \Big [ \frac{135}{8\pi^3\; n_{\rm eff}\; \gst\lst} \Big ]^{1/4} \; \Big ( 1-\frac{1}{2\alpha} \Big )^{1/4} \alpha^{-1/4} \; \mp
\ee
This temperature is the initial value for the classical cosmology after $\tr$. If we evolve the present $T_0= 2.7\; {\rm K}$
backward by means of the classical equations to the time when $H=\mp$, i.e. to $t=\tr$, we obtain a temperature of the order of 
$\mp$ and this value should coincide with (\ref{6.20}). Because of the very weak $\alpha$-dependence of $T(\tr)$, and assuming $n_{\rm eff} \gst\lst$
is of order unity, this is indeed the case for a wide range of $\alpha$-values, namely those for which $\OLS$ is not 
``anomalously close'' to $1$. 

Hence, contrary to many of the conventional models of inflation \cite{lily}, the RG cosmology
does not require a phase of ``reheating'' before the classical FRW evolution can (re-) start. 
The reason is clear:
Because of the energy transfer from the cosmological constant to the radiation there was a continuous ``heating'' during the NGFP epoch, as a 
consequence of which the temperature decreased only very slowly as $T\propto 1/t$ even though
$a\propto t^\alpha$ inflated rapidly with a large exponent $\alpha$ possibly.

We shall give a more quantitative description of the generation of density fluctuations elsewhere \cite{ourfluc}. 

\section{Discussion and conclusion}
\renewcommand{\theequation}{7.\arabic{equation}}
\setcounter{equation}{0}
In this paper we advocated the point of view that the scale dependence of the gravitational parameters 
has an impact on the
physics of the Universe we live in and we tried to identify known features of the 
Universe which could possibly be due
to this scale dependence. We discussed two possible candidates for such features: the entropy carried by  the radiation which fills the Universe
today, and a period of $\Lambda$-driven inflation directly after the big bang. 

As for the first point, we argued that within QEG the most likely RG trajectory is of type IIIa, 
predicting a positive cosmological constant whose magnitude decreases with scale. 
We saw that this leads to a continuous transfer
of energy from the vacuum to the matter sector. 
Already this process alone could  generate the entropy carried by the CMBR photons today. In this picture
the cosmological evolution started from a pure state; the entropy of the matter system
is caused by the ``coarse graining'' of the quantum gravitational dynamics which is forced
upon us because the optimal effective field theory $\Gamma_k [g_{\mu\nu}]$ changes as the
Universe expands. The time dependence of $k$ leads in particular to a time dependent
cosmological constant. It acts like a quintessence field \cite{cwquint,quint} in that 
is explains the present value of $\Lambda$ dynamically, its smallness being due to the 
Universe's old age. This quintessence field is a natural consequence of quantum field theory
and does not have to be introduced by hand. 

As for inflation, there is clearly no direct observational evidence for an inflationary epoch
in the early Universe which theory necessarily would have to explain. However, such an epoch would help
in understanding certain properties of the observed Universe, in particular as it can stretch
primordial density perturbations from sub- to super-Hubble scales. Allowing for an unrestricted energy
exchange between the vacuum and the matter sector we found solutions of the RG improved 
cosmological evolution equations with a phase of power law inflation immediately after the initial
singularity. In this phase $\Lambda$ dominates the matter energy density. 
The inflationary expansion gets  ``switched off'' automatically due to the RG running of $\Lambda(k)$.
For $ k \lesssim \mp$ the cosmology approaches that of a classical FRW model.
In this context the scale-, and hence time-, dependent cosmological constant plays the role of an inflaton 
which, again, does not need to be introduced by hand but rather arises as a quantum effect.

\vspace{10mm}
\noindent{\bf Acknowledgments}
M.~R. would like to thank the Astrophysical Observatory of Catania and   
the Albert Einstein Institute Potsdam for the hospitality extended
to him while this work was in progress.

\end{document}